\documentclass[amsmath,amssymb,amsbsy,reprint,prb,preprintnumbers,showpacs,superscriptaddress]{revtex4-2}
\usepackage{graphicx,color}
\usepackage{dcolumn}
\usepackage{bm}
\usepackage{braket}
\usepackage{mathtools}
\usepackage{ulem}
\usepackage[breaklinks,colorlinks=true,linkcolor=blue,urlcolor=blue,citecolor=blue]{hyperref}
\usepackage{times}
\usepackage{physics}
\usepackage{latexsym}
\usepackage{amsmath, amssymb}
\usepackage{mathtools}
\usepackage{multirow}

\newcommand{\be}{\begin{eqnarray}}
\newcommand{\ee}{\end{eqnarray}}

\begin{document}

\title{Production of lattice gauge-Higgs topological states in measurement-only quantum circuit}
\date{\today}
\author{Yoshihito Kuno} 
\affiliation{Graduate School of Engineering Science, Akita University, Akita 010-8502, Japan}
\author{Ikuo Ichinose} 
\thanks{A professor emeritus}
\affiliation{Department of Applied Physics, Nagoya Institute of Technology, Nagoya, 466-8555, Japan}


\begin{abstract}
By imaginary-time evolution with Hamiltonian, an arbitrary state arrives in the system's ground state. In this work, we conjecture that this dynamics can be simulated by measurement-only circuit (MoC), where each projective measurement is set in a suitable way.
Based on terms in the Hamiltonian and ratios of their parameters (coefficients), we propose a guiding principle for the choice of the measured operators called stabilizers and also the probability of projective measurement in the MoC.  
In order to examine and verify this conjecture of the parameter ratio and probability ratio correspondence in a practical way, we study a generalized (1+1)-dimensional $Z_2$ lattice gauge-Higgs model, whose phase diagram is very rich including symmetry-protected topological phase, deconfinement phase, etc. 
We find that the MoC constructed by the guiding principle reproduces phase diagram very similar to that of the ground state of the gauge-Higgs Hamiltonian. The present work indicates that the MoC can be broadly used to produce interesting phases of matter, which are difficult to be simulated by ordinary Hamiltonian systems composed of stabilizer-type terms.
\end{abstract}


\maketitle
\section{Introduction}
Measurement of quantum many-body system induces nontrivial dynamical effects and produces exotic phases of matter. 
One of the most interesting phenomena induced by measurements is entanglement phase transition in hybrid random unitary circuits \cite{Li2018,Skinner2019,Li2019,Vasseur2019,Chan2019,Szyniszewski2019,Choi2020,Bao2020,Jian2020,Zabalo2020,Sang2021,Sang2021_v2,Nahum2021,Sharma2022,Fisher2022_rev}. 
This phase transition phenomenon emerges in various hybrid circuits including time-evolution operator by many-body Hamiltonian \cite{Fuji2020,Goto2020,Tang2020,Lunt2020,Turkeshi2021,Kells2022,Fleckenstein2022,KOH2022}.  
High entanglement of states generated by unitary time evolution is suppressed by the measurements. 
Also, as typical non-equilibrium dynamics, the spread of entanglement and scrambling of quantum information are suppressed. 
Without time-evolution unitary, measurement-only quantum circuit (MoC) \cite{Lang2020,Ippoliti2021} also displays striking phenomena, i.e., combination of multiple kinds of measurements, some of which are not commutative with each other, can induce novel phase transitions and generate non-trivial states such as measurement-only thermal state without exhibiting area law of entanglement entropy \cite{Ippoliti2021}, symmetry protected topological (SPT) state \cite{Lavasani2021,Klocke2022} and topological order \cite{Lavasani2021_2}.
It should be remarked that these phase transitions in the MoC exhibit some universal behavior at transition points as reported in recent studies \cite{Ippoliti2021,Lavasani2021,Lavasani2021_2,Klocke2022}.

In the previous works  \cite{Lavasani2021,Lavasani2021_2,Klocke2022}, sequential stabilizer projective measurements are operated to the system as a MoC and emergence of non-trivial states is observed.
There, interestingly enough, the resultant phase diagram of the MoC is similar and almost identical to that of the ground state of the Hamiltonian composed of the operated stabilizers. 
For example, the phase diagram of the cluster spin Hamiltonian with local $X_j$ terms \cite{Zeng2016} can be reproduced in the MoC by varying the probability ratio of projective measurements between the cluster-spin and the local $X_j$ operators \cite{Lavasani2021}. 
This result implies that the coefficient ratio between competing terms in the stabilizer Hamiltonian corresponds to the probability ratio between the projective measurements of the stabilizers, which 
anti-commute with each other in the MoC. 
In Ref.~\cite{Klocke2022}, an interesting conjecture is mentioned that the steady state in the MoC including stabilizer measurements is close to the ground state obtained by an imaginary-time evolution of the corresponding stabilizer Hamiltonian. 
However, details of the above interesting conjecture have not been studied yet, and further concrete examples (both analytical and numerical ones) clarifying the correspondence are still lacking. 

In this work, we focus on a Hamiltonian composed of stabilizer-type terms, some of which are anti-commutative with each other (shown in Eq.~(\ref{stab_H})), and a corresponding MoC and the process of its numerical simulation. 
We shall study the following two subjects to clarify the above conjecture: 
\begin{enumerate}
\item Based on the qualitative conjecture of parameter ratio-probability ratio correspondence (PRC) suggested in \cite{Lavasani2021,Klocke2022}, we investigate the PRC in a qualitative level by using the imaginary-time path integral formalism and the MoC of the Gottesman-Knill stabilizer simulation \cite{Gottesman1997,Aaronson2004,Nielsen_Chuang}. Comparing the path-integral formalism and the MoC, 
we strengthen the conjecture. Some simple analytical examples are also shown. 
Although rigorous mathematical proof for the PRC is not given in this work, our study supports the conjecture in a substantial way.
\item To investigate the PRC concretely, we study an interesting system of great physical significance in high-energy physics and also condensed matter physics. 
That is, we focus on a (1+1)-D $Z_2$ lattice gauge-Higgs model. Recently, the Higgs phase of the lattice gauge theory (LGT) \cite{Fradkin1979,Kogut1979} is suggested to have properties of the symmetry-protected topological (SPT) phase \cite{Verresen2022} and also the ground state phase diagram of the gauge-Higgs Hamiltonian was studied in \cite{Borla2021,Verresen2022}. 
\item Instead of working on the Hamiltonian system of the above model, we numerically study its phase diagram of mixed state by applying the mixed-state update methods of stabilizer dynamics employed in \cite{Gullans2020,Ippoliti2021} to examine the PRC. 
That is, MoC corresponding to the gauge-Higgs Hamiltonian is constructed by using the guiding principle of the PRC.
We draw the mixed-state phase diagram of the MoC and find its clear correspondence to the LGT Hamiltonian system. 
This also indicates that the MoC with suitable stabilizer measurements produces interesting gauge-theoretical states predicted as a ground state of LGTs. 
We further study phase transition criticality for some typical parameter sweeps by finite-size scaling (FSS) analysis. We comment on the critical exponents obtained via the MoC.
\end{enumerate}

The rest of this paper is organized as follows. 
In Sec.~II, we shall discuss the PRC conjecture in a qualitative level. We show simple concrete examples for examination of the PRC, and discuss the extension of the PRC to the mixed-state case.
In Sec.~III, we shall introduce the Hamiltonian of (1+1)-D $Z_2$ lattice gauge-Higgs model and shortly review its ground state properties. Then, we introduce the setup of the MoC for searching the properties of the ground state of the gauge-theory Hamiltonian rather in detail. 
There, the PRC plays a role of the guiding principle.
In Sec.~IV, we show the results of the numerical study of the MoC corresponding to the (1+1)-D $Z_2$ lattice gauge-Higgs model. Detailed discussions on the numerical results and study of phase transition criticality are given. 
Section IV is devoted to conclusion.


\section{Conjecture of parameter ratio and probability ratio correspondence}
\begin{figure}[t]
\begin{center} 
\includegraphics[width=8cm]{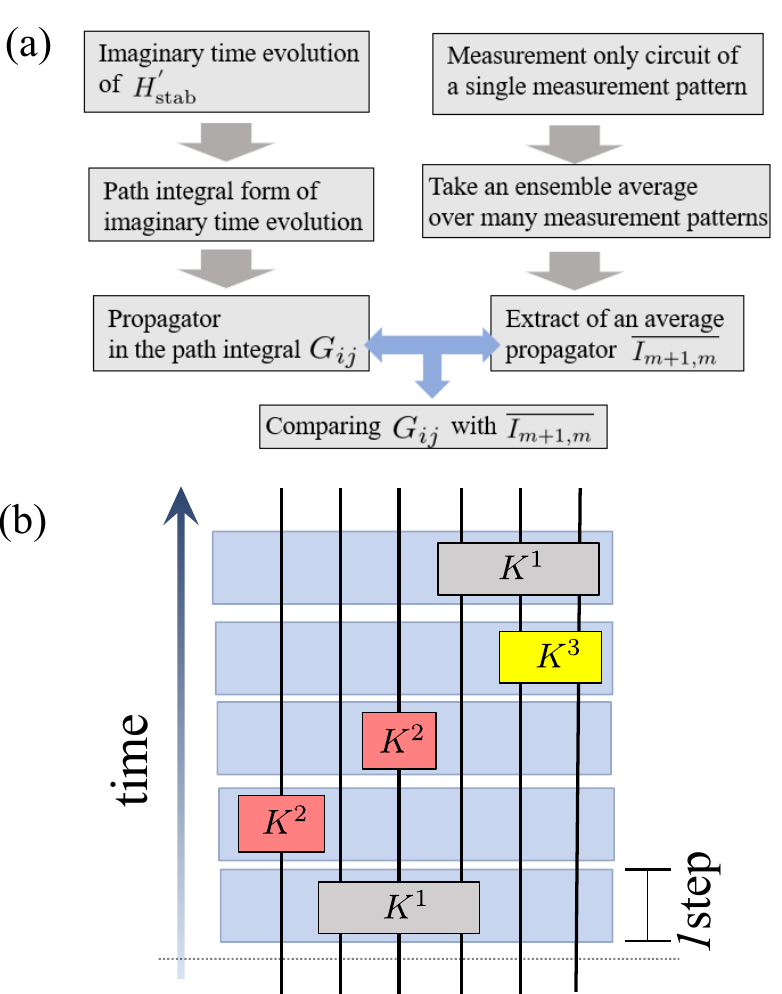}  
\end{center} 
\caption{
(a) Flowchart of comparison between the imaginary-time evolution of $H^{'}_{\rm stab}$ and the measurement-only circuit. 
(b) Schematic image of time evolution of the MoC. 
As an example, three different types of stabilizers $K^{\alpha}_j$ ($\alpha=1,2,3$) are considered. 
During a single time step, one of three kinds of the stabilizer is chosen with the probability $p^\alpha$ and its projective measurement is carried out, where $\sum^3_{\alpha=1}p^\alpha=1$ and the position (site) of the measurement is chosen randomly with equal probability.}
\label{Fig1}
\end{figure}

In this section, we start with a random-coupling Hamiltonian, each term of which is a stabilizer. 
This type of Hamiltonian would be expected to have a corresponding counterpart MoC. 
That is, the both systems share a very close ground state phase diagram, here `ground state' of the MoC means steady states appearing after a long time evolution. 
In order to examine the conjecture, we first introduce the imaginary-time evolution and its path integral formalism. Second we explain the setup of the corresponding MoC and consider an ensemble of steady states obtained by the time evolution of the MoC. 
Even though the expression of the ensemble of steady states and the time-evolution propagator are mathematically not rigorous, their descriptions are useful to compare the MoC and the imaginary-time evolution of the Hamiltonian system. 
In fact, we obtain a useful insight for the PRC. The flowchart of this section is shown in Fig.~\ref{Fig1} (a). 

\subsection{Considered Hamiltonian}
We start to consider a general binary random-coupling stabilizer Hamiltonian in one dimension defined as follows,
\begin{eqnarray}
H_{\rm stab}=\sum^{L-1}_{j=0}\sum^{M}_{\alpha=1}J^{\alpha}_{j}K^{\alpha}_{j},
\label{stab_H}
\end{eqnarray}
where $L$ is a total number of sites $\{ j\}$,  $\alpha$ represents $M$-types of stabilizers anti-commuting with each other, i.e., 
$\{ K^{\alpha}_j\}$ satisfy $[K^{\alpha}_j,K^{\alpha}_k]=0$ and $(K^{\alpha}_j)^2=1$,
and for different types of stabilizers, 
$[K^{\alpha}_j,K^{\beta}_k]\neq 0$ and $\{K^{\alpha}_j,K^{\beta}_k\}=0$ ($\alpha\neq \beta$) \cite{stab}.
The couplings are local and binary for $\forall j$, $J^{\alpha}_{j}=\pm J^{\alpha}$, $J^{\alpha}>0$. 
The arguments throughout this work apply only to the type of the Hamiltonian $H_{\rm stab}$. 
In general, the model has a rich ground-state phase diagram depending on the choice of the stabilizers and exhibits clear phase transitions on varying values of parameters. Note that the ground state is not generally unique, depending on the number of stabilizers.

\subsection{General setup of measurement-only circuit}
We construct a MoC counterpart to the above stabilizer Hamiltonian $H_{\rm stab}$ by using the PRC as guiding principle. 
In the MoC, we choose a single stabilizer among the different types of $K^\alpha_{j_{0}}$ with a probability $p^{\alpha}$ and choose a target site $j_0$ with equal probability $1/L$ at each time step. 
We set the probability condition of the choice of the type of the stabilizer, such as $\sum_{\alpha}p^\alpha=1$. 
The setup is the same with that employed in the previous works \cite{Lavasani2021,Klocke2022}. 
Then, we perform the projective measurement corresponding to the stabilizer $K^\alpha_{j_{0}}$. 
We expect that after large number of time steps, a state reaches a steady state for most of cases. If $p^{\alpha'}$ with a particular $\alpha'$ is dominant, the steady state is stabilized by $K^{\alpha'}_j$, corresponding to a stabilizer state. 
Schematic example of $M=3$ case of the MoC (three different types of stabilizer projective measurements) is shown in Fig.~\ref{Fig1} (b). 

\subsection{Parameter fixing of $H_{stab}$ and simplified MoC}
In this work, we simulate MoCs (numerically) by employing simplified stabilizer circuits \cite{Gottesman1997,Aaronson2004}, in which the information of the overall sign of the observed value of the stabilizers by projective measurements is not stored as in many other previous studies \cite{Nahum2017,Lunt2020,Sang2021,Turkeshi2021,Lavasani2021,Klocke2022}. 
In the practical simulation, we fix the measured value to $+1$ for all stabilizers in the MoC. In other words, this fixing  means that a projective measurement of stabilizes denoted by $P^{\alpha}_j$ is fixed as $P^{\alpha}_j=\frac{1+K^{\alpha}_j}{2}$ at each time step. Corresponding to this setup of MoCs, the following Hamiltonian, instead of $H_{\rm stab}$, to be considered for clarifying the subsequent discussion,
\begin{eqnarray}
H'_{\rm stab}=-\sum^{L-1}_{j=0}\sum^{M}_{\alpha=1}J^{\alpha}K^{\alpha}_{j},
\label{Hstab2}
\end{eqnarray}
where the binary random couplings have been set as $J^{\alpha}_j\to -J^{\alpha}$. 

Previous studies \cite{Lavasani2021,Klocke2022} showed that the ground state phase diagram of $H_{\rm stab}$ or $H'_{\rm stab}$ is significantly close to that of the MoC, which is determined by an ensemble average of the measurement pattern of the MoC. This fact implies that the ratio of parameters $\frac{J^\alpha}{J^{\beta}}$ is related to the ratio of probabilities $\frac{p^{\alpha}}{p^{\beta}}$, that is, $\frac{J^\alpha}{J^{\beta}}\longleftrightarrow\frac{p^{\alpha}}{p^{\beta}}$. 
This relation is nothing but the explicit form of ``parameter ratio-probability ratio correspondence''. 
In what follows, we study the conjecture of the PRC in a qualitative level by employing imaginary-time path integral and by focusing on the averaged states in the MoC. 
We further examine the PRC for small size systems as a concrete example.

\subsection{Imaginary-time evolution}
For the stabilizer Hamiltonian $H'_{\rm stab}$, the ground state can be generated by imaginary-time evolution, which is used by various numerical simulations, such as path-integral quantum Monte-Carlo method \cite{Avella2013}. 
If degenerate ground states exist in Hamiltonian, we expect that the ground state generated by the imaginary-time evolution is one of  linear combinations of them (that is, a pure state).
Especially for a spontaneously-symmetry-broken phase, one of the ground states with a definite order parameter is to be chosen.

The imaginary-time evolution starting with a state $|\psi(0)\rangle$ generates a final state as 
\begin{eqnarray}
|\psi(\tau)\rangle =e^{-\tau H'_{\rm stab}}|\psi(0)\rangle,
\end{eqnarray}
where $\tau$ is the imaginary-time interval (regarded as inverse temperature). For sufficiently large $\tau$, we assume the final state $|\psi(\tau)\rangle$ reaches the ground state of the $H'_{\rm stab}$. 
We split the interval $\tau$ into $N$segments ($N\gg 1$) and insert identities composed of a complete set of basis,   
\begin{eqnarray}
|\psi(\tau)\rangle &=&\sum_{\{\ell\}}
|\ell_{N}\rangle \langle \ell_{N}|e^{-\delta \tau H'_{\rm stab}}|\ell_{N-1}\rangle\nonumber\\
&&\cdots \langle \ell_{1}|e^{-\delta \tau H'_{\rm stab}}|\ell_{0}\rangle \langle \ell_{0}|\psi(0)\rangle\nonumber\\
&=&\sum_{\ell_{N}}|\ell_{N}\rangle \biggl[\sum_{{\bf \{\ell\}}-\ell_N}\prod^{N-1}_{j=0}G_{j+1,j}\langle \ell_{0}|\psi(0)\rangle\biggr],\\
G_{i,j}&\equiv& \langle \ell_{i}|e^{-\delta\tau H'_{\rm stab}}|\ell_j\rangle.
\end{eqnarray}
where $\delta \tau=\tau/N$, $\sum_{\ell}|\ell\rangle\langle\ell|=1$, i.e., $\{|\ell\rangle\}$ is a set of basis and we have employed Suzuki-Trotter decomposition \cite{Suzuki1976}, and $G_{j+1,j}$ is a propagator for small discrete time step $\delta \tau$. 
The above is a discrete imaginary-time path integral, and
the imaginary-time dynamics is governed by the propagator $G_{j+1,j}$.

\subsection{Time evolution of MoC and ensemble state}
As the next step, we turn to the MoC starting with a state $|\psi(0)\rangle$, where a sufficient large number of discrete time steps denoted by $t_N$ are performed. In the MoC, a measurement pattern of time evolution is selected (called unraveling and this is a single stochastic process). 
Then, we assume that the final state reaches a steady state. This state can be written by \cite{Gullans2020} 
\begin{eqnarray}
|\psi(t_N)_{{\vec \alpha},{\vec j}}\rangle &=& C_{{\vec \alpha},{\vec j}} Q_{{\vec \alpha},{\vec j}} |\psi(0)\rangle,\\
Q_{{\vec \alpha},{\vec j}}&=&P^{\alpha_{N}}_{j_N}P^{\alpha_{N-1}}_{j_{N-1}}P^{\alpha_{N-2}}_{j_{N-2}}\cdots P^{\alpha_{1}}_{j_1}, \\
P^{\alpha_{m}}_{j_m}&=&\frac{1}{2}(1+ K^{\alpha_{m}}_{j_m}).
\end{eqnarray}
Here the single measurement pattern is represented by labels ${\vec \alpha}$ and ${\vec j}$, where ${\vec \alpha}=(\alpha_1, \alpha_2, \cdots, \alpha_{t_N})$, $\alpha_m (=1,\cdots, M)$ represents the type of the stabilizer at $m$-th time step with a probability $p^{\alpha}$, and ${\vec j}=(j_1, j_2, \cdots, j_{t_N})$, $j_m(=0,\cdots, L-1)$ represents the position of the performed  projective measurement at $m$-th time step. $P^{\alpha_m}_{j_m}$ is $\alpha$-types projective measurement at $m$-th time step. 
$C_{{\vec \alpha},{\vec j}}$ is a normalization constant of the state, which depends on the single measurement pattern $({\vec \alpha},{\vec j})$. 

As in the imaginary-time evolution in the above, we insert many identities composed of a complete set of basis between neighboring projective operators,  
\begin{eqnarray}
|\psi(t_N)_{{\vec \alpha},{\vec j}}\rangle &\propto&\sum_{{\bf \ell}}
|\ell_{N}\rangle \langle \ell_{N}|P^{\alpha_{N}}_{j_N}|\ell_{N-1}\rangle\nonumber\\
&&\times
\langle \ell_{N-1}|P^{\alpha_{N-1}}_{j_{N-1}}|\ell_{N-2}\rangle
\cdots 
\langle \ell_{1}|P^{\alpha_{0}}_{j_{0}}|\ell_{0}\rangle \langle \ell_{0}|\psi(0)\rangle\nonumber\\
&=&\sum_{\ell_{N}}|\ell_{N}\rangle \biggl[\sum_{{\bf \{\ell\}}-\ell_{N}}\prod^{N-1}_{m=0}I_{m+1,m}\langle \ell_{0}|\psi(0)\rangle\biggr],\\
I_{m,m-1}&\equiv& \langle \ell_{m}|P^{\alpha_{m}}_{j_{m}}|\ell_{m-1}\rangle,
\end{eqnarray}
where we used $\sum_{\ell}|\ell\rangle\langle\ell|=1$ and  $\{|\ell\rangle\}$ is the same set of basis used in the imaginary-time evolution in the above.
In what follows, we shall ignore normalization factor of the sequence of the projective measurements.

To compare the above state in the MoC with the ground state created by the imaginary-time evolution $|\psi(\tau)\rangle$, we further proceed with the above consideration of the MoC. 
We take ensemble average over many single measurement patterns creating a steady state $|\psi(t_N)\rangle$. 
Here, we introduce sample label $s$. 
Each sample of measurement pattern is labeled as 
$({\vec \alpha},{\vec j}) \to ({\vec \alpha}^s,{\vec j}^s)$, where $s=1,2,\cdots, N_p$ and $N_p$ is the total number of the samples. 
Then, an averaged steady state is described by
\begin{eqnarray}
&&\overline{|\psi(t_N)\rangle} = \frac{1}{N_p}\sum_{s}C_{{\vec \alpha}^s,{\vec j}^s} Q_{{\vec \alpha}^s,{\vec j}^s} |\psi(0)\rangle\nonumber\\
&&\propto \frac{1}{N_p}\sum_{s} Q_{{\vec \alpha}^s,{\vec j}^s} |\psi(0)\rangle\nonumber\\
&&=\frac{1}{N_p}\sum_{s} \sum_{\ell_{N}}|\ell_{N}\rangle \biggl[\sum_{{\bf \{\ell\}}\neq \ell_{N}}\prod^{N-1}_{m=0}I^s_{m+1,m}\langle \ell_{0}|\psi(0)\rangle\biggr]\nonumber\\
&&\stackrel{N_p\to\infty}{=}\sum_{\ell_{N}}|\ell_{N}\rangle \biggl[\sum_{{\bf \{\ell\}}- \ell_{N}}\prod^{N-1}_{m=0}\overline{I_{m+1,m}}\langle \ell_{0}|\psi(0)\rangle\biggr]\nonumber\\
&&\equiv \overline{Q(t_N)} |\psi(0)\rangle, \label{Qave}
\end{eqnarray}
where 
\begin{eqnarray}
\overline{I_{m+1,m}}\equiv \langle \ell_{m+1}|\biggl(\sum_{\alpha,j}\frac{p^{\alpha}}{L}P^{\alpha}_{j}\biggl)|\ell_{m}\rangle.
\end{eqnarray}
From this form, the averaged state $\overline{|\psi(t_N)\rangle}$ is approximately determined by an ensemble averaged propagator $\overline{I_{m+1,m}}$. 
Note that we here ignore the factor $C_{{\vec \alpha}^s,{\vec j}^s}$.

\subsection{Qualitative relationship of propagators and concrete relation between parameter ratio and probability ratio} 
Now we compare the two propagators $G_{j+1,j}$ and $\overline{I_{m+1,m}}$ to obtain a relation between them. 
It is expected that if the structure of these propagators is close, the obtained ground state $|\psi(\tau)\rangle$ must be close to the steady state $\overline{|\psi(t_N)\rangle}$. 
That is, 
\begin{eqnarray}
G_{j+1,j}\longleftrightarrow\overline{I_{m+1,m}}
\Longleftrightarrow |\psi(\tau)\rangle \longleftrightarrow \overline{|\psi(t_N)\rangle},
\end{eqnarray}
Here, $\longleftrightarrow$ means ``close structure''.

From the above observation, if $|\psi(\tau)\rangle \longleftrightarrow \overline{|\psi(t_N)\rangle}$ is correct, we can conclude 
$G_{j+1,j}\longleftrightarrow\overline{I_{m+1,m}}$. 
Then, by comparing the internal structure of the matrices $G_{j+1,j}$ with that of $\overline{I_{m+1,m}}$, we can obtain important insight and relationship between model parameters of $H'_{\rm stab} $ [$\{ J^{\alpha}\}$] and emergent probability   $\{ p^{\alpha}\}$ and types of stabilizers in the corresponding MoC.

In general, it is difficult to find strict and rigorous relations between model parameters of $H'_{\rm stab}$  [$\{ J^{\alpha}\}$] and probability $\{ p^{\alpha}\}$ for many-body system due to large Hilbert space dimension and large dimension of the matrices of propagator. 
However, we can find a qualitative relation if we consider a simple Hamiltonian and its corresponding MoC. 
We study two concrete examples given as follows: 

\uline{Case (I)}: 
Single spin Hamiltonian,  
$$
H^{'}_{{\rm stab} 1}=-J_1Z-J_2X,
$$
where $Z$ and $X$ are Pauli operators of  single $1/2$-spin, $J_{1(2)}>0$. 
Note that $Z$ and $X$ are different types of stabilizers, which are anti-commutative with each other, corresponding to $M=2$ and $L=1$ case in Eq.~(\ref{Hstab2}). 
For this Hamiltonian, the propagator of the imaginary-time path integral is $G_{j+1,j}=\langle \ell_{j+1}|e^{-\delta\tau H^{'}_{\rm stab 1}}|\ell_j\rangle$, where the set of basis is $\{|\ell_j\rangle\}=\{|\uparrow\rangle, \:\:|\downarrow\rangle \}$ where $Z|\uparrow\rangle=|\uparrow\rangle$ and $Z|\downarrow\rangle=-|\downarrow\rangle$. 
The matrix form of $G_{j+1,j}$ is obtained by the practical calculation as
\begin{eqnarray}
(G_{ij})\approx
\begin{bmatrix}
   e^{\delta\tau J_1}\cosh(\delta \tau J_2) & e^{-\delta\tau J_1}\sinh(\delta \tau J_2) \\
   e^{\delta\tau J_1}\sinh(\delta \tau J_2)
   & e^{-\delta\tau J_1}\cosh(\delta \tau J_2)
\end{bmatrix},
\end{eqnarray}
where we have ignored the contribution from the commutators of $X$ and $Z$ (due to $\delta \tau \ll 1$).

Let us turn to the propagator of the MoC. 
The MoC corresponding to $H^{'}_{{\rm stab} 1}$ includes a single site projective measurement of $Z$ and $X$ with probability $p^A$ and $p^B$, respectively, where $p^A+p^B=1$. 
Then, the matrix form of the averaged propagator $(\overline{I_{ij}})$ is given by 
\begin{eqnarray}
(\overline{I_{m+1,m}})= 
\begin{bmatrix}
   \frac{1+p^A}{2} & \frac{p^B}{2} \\
   \frac{p^B}{2}
   & \frac{1-p^A}{2}
\end{bmatrix}.
\end{eqnarray}

We compare the components of the two matrices $(G_{j+1,j})$ and $(\overline{I_{m+1,m}})$. 
The following four relations are then obtained (we ignoring an overall factor $e^{i\rho}$), 
\begin{eqnarray}
\mbox{1 column:}\:&& e^{\delta\tau J_1}\cosh(\delta \tau J_2) \longleftrightarrow \frac{1+p^A}{2}, \nonumber\\
&&e^{\delta\tau J_1}\sinh(\delta \tau J_2)\longleftrightarrow \frac{p^B}{2}.\\
\mbox{2 column:}\:&& e^{-\delta\tau J_1}\sinh(\delta \tau J_2) \longleftrightarrow \frac{p^B}{2}, \nonumber\\ 
&&e^{-\delta\tau J_1}\cosh(\delta \tau J_2)
\longleftrightarrow \frac{1-p^A}{2}.
\end{eqnarray}
At first glance, we note that an increase (decrease) of the ratio $J_1/J_2$ corresponds to an increase (decrease) of $p^A/p^B$. 
More precisely for $\delta \tau \ll 1$, we expand each component up to $\mathcal{O}(\delta \tau)$, then we reach the following relations,
\begin{eqnarray}
\mbox{1 column:}\: 
1+\delta\tau J_1 &\longleftrightarrow& \frac{1+p_A}{2}, \nonumber\\ 
\delta\tau J_2  &\longleftrightarrow& \frac{p_B}{2}.\\
\mbox{2 column:}\: 
\delta \tau J_2 &\longleftrightarrow& \frac{p_B}{2}, \nonumber\\ 1-\delta\tau J_1 &\longleftrightarrow& \frac{1-p_A}{2}.
\end{eqnarray}
By requiring $G_{j+1,j}=C_0(\overline{I_{m+1,m}})$, the comparing $(1,1)$-component with $(2,2)$-component leads to $\delta\tau J_1=\frac{C_0}{2}p^A$, and also the comparing $(1,2)$-component with $(2,1)$-component leads to
$\delta\tau J_2=\frac{C_0}{2}p^B$.
Thus, we obtain 
\begin{eqnarray}
\frac{J_1}{J_2} \longleftrightarrow \frac{p^A}{p^B}. 
\end{eqnarray}
This is a concrete form of the PRC between the imaginary-time path integral formalism of $H'_{\rm stab1}$ and its corresponding counterpart MoC. 

\uline{Case (II)}: 
As second example, we consider a three-site cluster spin model, 
\begin{eqnarray}
H^{'}_{\rm stab 2}=\sum^{2}_{j=0}[-J_1Z_{j-1}X_jZ_{j+1}-J_2X_j],
\end{eqnarray}
where periodic boundary conditions are imposed and $J_{1(2)}>0$. 
The operators $ZXZ$ and $X$ are different types of stabilizers, which are commutative/anti-commutative with each other depending on their locations, and the model corresponds to the $M=2$ and $L=3$ case in Eq.~(\ref{Hstab2}).

We first consider the matrix propagator of the imaginary-time propagation $\langle \ell|e^{-\delta\tau H^{'}_{\rm stab 2}}|\ell'\rangle= G_{j+1,j}$. Here, we employ eigenstates of $\{ X_j\}$ as a complete set of basis, and therefore,
\begin{eqnarray}
\{|\ell\rangle\}&=&\{|+++\rangle, \:\:|-++\rangle, \:\:
|+-+\rangle, \:\:|++-\rangle,\:\:\nonumber\\
&&|--+\rangle, \:\:|-+-\rangle,\:\:
|+--\rangle, \:\:|---\rangle
\}.
\end{eqnarray}
$G_{j+1,j}$ is $8\times 8$ matrix. Components of the propagators are approximately obtained by ignoring the contributions from the stabilizers' commutators by the Suzuki-Trotter decomposition. The $8\times 8$ full matrix is explicitly shown in Appendix A. 

We turn to the propagator of the MoC. 
By the PRC guiding principle,
the MoC corresponding to $H^{'}_{{\rm stab} 2}$ includes a single projective measurement of $ZXZ$ and $X$ with probability $p^A$ and $p^B$ at a single time step, respectively. Here, $p^A+p^B=1$ and the measurement site is chosen randomly with equal probability $1/L=1/3$. 
Then, the $8\times 8$ full matrix of the averaged propagator $(\overline{I_{m+1,mj}})$ is also directly calculable. 
The full form is also shown in Appendix A. 

Now, we employ the same strategy to the case I. That is, we compare the components of the two matrices $(G_{j+1,j})$ and $(\overline{I_{m+1,m}})$. 
Fortunately, we find only five relations given by (the detailed calculation is shown in Appendix A) 
\begin{eqnarray}
 1+3\delta \tau J_2 &\longleftrightarrow& p^A/2+p^B, \nonumber\\
 1+\delta \tau J_2 &\longleftrightarrow& p^A/2+2p^B/3, \nonumber\\
 1-\delta \tau J_2 &\longleftrightarrow& p^A/2+p^B/3, \nonumber\\
 1-3\delta \tau J_2 &\longleftrightarrow& p^A/2,\nonumber\\ 
\delta\tau J_1&\longleftrightarrow& p^A/6.
\end{eqnarray}
From the above relations, if we require  $G_{j+1,j}=C_0(\overline{I_{m+1,m}})$, then $\delta\tau J_1=\frac{C_0}{6}p^A$, and the relations $1+3\delta \tau J_2 = C_0(p^A/2+p^B)$ and $1+\delta \tau J_2 = C_0(p^A/2+2p^B/3)$ lead to $\delta \tau J_2=\frac{C_0}{6}p^B$. 
Hence, we have
\begin{eqnarray}
\frac{J_1}{J_2} \longleftrightarrow \frac{p^A}{p^B}.
\end{eqnarray}
This relation is the same with that obtained in the case I.

We showed that the genuine PRC relation appears for the above two concrete cases classified in the type of Hamiltonian of Eq.~(\ref{stab_H}) by comparing the imaginary-time path integral formalism and MoC.

This is the genuine PRC relation between the imaginary-time path integral formalism and the MoC.

\subsection{PRC for mixed-state dynamics}
We have strengthened the PRC conjecture in previous subsections, following
the previous works for pure-state evolution, which  numerically imply the PRC in some parts of phase diagrams of certain models \cite{Lavasani2021,Klocke2022}.

Forwarding the discussion one step further, we extend the above discussion on the pure state to the mixed state, in particular, starting with an infinite-temperature mixed state. 
Under the imaginary-time evolution, the density matrix dynamics is given by
\begin{eqnarray}
\rho(\tau)=e^{-\tau H} \rho(0)e^{\tau H},
\end{eqnarray}
where $H$ is a Hamiltonian and a suitable normalization of $\rho(\tau)$ is assumed.
Here, we set $\rho(0)$ to an infinite-temperature state. 
We expect that this approach is efficient to detect a degenerate ground-state multiplet of the system, and there, the steady mixed state can be constructed by the ground-state multiplet. 
In fact, for sufficient large $\tau$, the state $\rho(\tau)$ results in a ground state, which is a multiplet if the ground state of $H$ is degenerate.

Similar observation with the above can be applied to the MoC for each single measurement pattern. 
We consider the ensemble average of density matrix averaged over samples of measurement patterns. 
If we employ the averaged time-evolution operator of the MoC $\overline{Q(t_N)}$ in Eq.(\ref{Qave}), the averaged time-evolved density matrix $\overline{\rho(t_N)}$ is approximately given by
\begin{eqnarray}
\overline{\rho(t_N)}\sim \overline{Q(t_N)} \rho(0)\overline{Q(t_N)}^{\dagger}.
\end{eqnarray}
One might expect that the PRC, similar to the pure-state system, holds for the above mixed-state system 
since the propagators for the update are the same. 

However for the mixed state in the quantum circuit, $\overline{\rho(t_N)}$ is \textit{not} commonly used for calculation of physical quantities such as entropy, entanglement entropy, etc \cite{Sharma2022}. 
More precisely in the MoC, physical quantities are obtained for each single measurement path and so-obtained results are averaged over various measurement patterns.
Then rigorously, it is a nontrivial question whether the PRC holds for the MoC of mixed states from the view point of quantum-mechanical coherence. Therefore, it is very important to examine if the PRC holds for the mixed states in the MoC and to show its concrete examples. 
In this work, we address this problem by employing numerical methods. We shall study MoCs for a lattice gauge models classified in the type of Hamiltonian of Eq.~(\ref{stab_H})  as a concrete example, which has a very rich and interesting phase diagram. 
Sometimes, degenerate ground states emerge, hence for practical use, we employ the mixed state update.

\begin{figure}[t]
\begin{center} 
\includegraphics[width=8.5cm]{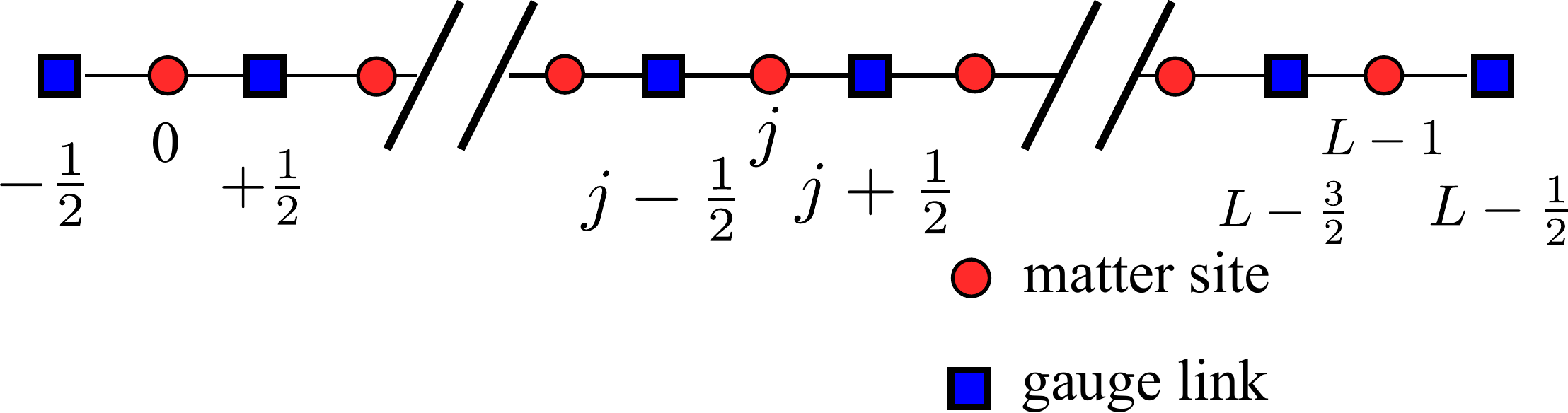}  
\end{center} 
\caption{Lattice-link setting for extended cluster model. Open boundary conditions are imposed. 
The total number of the matter site is $L$ and the total number of the gauge link is $L+1$.
}
\label{Fig2}
\end{figure}
\section{(1+1)-D $Z_2$ lattice gauge-Higgs model and its measurement-only circuit counterpart}
In previous section, we discussed the PRC between the imaginary-time formalism of Hamiltonian and the ensemble average of the MoC. Certain simple examples were investigated there. 
We shall further examine and strengthen this guiding principle by investigating another model of great physical interest. In this section, we study a lattice gauge model called  ``(1+1)-D $Z_2$ lattice gauge-Higgs model'' with open boundary conditions. 
Its global ground state phase diagram was recently studied rather in detail\cite{Borla2021,Verresen2022}. The model includes interesting phases, and the study on it reveals an important relationship between gauge theory and topological order in condensed matter. 

We address the following issue: 
Based on the PRC guiding principle, whether or not a suitably-chosen MoC generates steady states phase diagram of which is similar or identical to the ground state phase diagram of the target gauge-theory model. 
We shall present a suitable setup of the MoC and clarify this issue.

\subsection{Model Hamiltonian proposed in Refs. \cite{Borla2021,Verresen2022}}

We introduce a gauge-lattice as shown in Fig.~\ref{Fig2}, where spin-$1/2$ degrees of freedom reside both on matter sites and gauge links in one spatial dimension. 
Therefore, the total degrees of freedom are $L_t\equiv 2L+1$ spins.
We focus on the following cluster spin Hamiltonian \cite{Borla2021,Verresen2022},
\begin{eqnarray}
H_{Z_2}&=&\sum^{L-1}_{j=0}\biggl[-K_1\sigma^{x}_{j-1/2}X_{j}\sigma^{x}_{j+1/2}-K_2\sigma^z_{j+1/2}\biggl]\nonumber\\
&+&\sum^{L-1}_{j=0}\biggl[-J_1Z_{j}\sigma^{z}_{j+1/2}Z_{j+1}
-J_2X_{j}\biggl],
\label{HZ2}
\end{eqnarray}
where $X_j$ and $Z_j$ are Pauli operators defined on matter sites and $\sigma^x_{j\pm\frac{1}{2}}$ and $\sigma^z_{j\pm\frac{1}{2}}$ are also Pauli operators on gauge links. 
We consider open boundary conditions throughout this work. 
Note that the boundaries of the system are the gauge links as shown in Fig.~\ref{Fig2}. 
The model has two important symmetries: (I) Parity symmetry, $P\equiv \prod^{L-1}_{j=0}X_{j}$ and (II) Magnetic symmetry $W\equiv \prod^{L-1}_{j'=-1}\sigma^z_{j'+\frac{1}{2}}$, resulting in $Z_2\times Z_2$ symmetry, which has been referred as key symmetry for SPT phase \cite{Son2011,Son2012,Bahri2015,Verrsen2017}.  

The model $H_{Z_2}$ in (\ref{HZ2}) reduces to well-known (1+1)-D $Z_2$ lattice gauge-Higgs for $K_1/K_2\to \infty$, and also it has SPT properties of condensed-matter physics in certain parameter region. 
More precisely from the gauge-theoretical point of view, the $K_1$-term acts as energetic penalty caused by breaking of  Gauss' law constraint. 
On the other hand, the $K_2$-term hinders fluctuations of the gauge field.
$J_1$-term is a cluster term, interpreted as a matter-($Z_2$)gauge coupling and also it is a topological stabilizer protected by $Z_2\times Z_2$ symmetry in SPT literature, and $J_2$-term acts as a chemical potential of the matter and also is regarded as a `transverse field' competing with the cluster term. 
The above four terms are different types of stabilizers from the MoC point of view. 

The ground state of $H_{Z_2}$ and its phase diagram were studied in detail \cite{Borla2021,Verresen2022}. 
The model has four ground state phases: (1) Higgs=SPT phase, 
(2) deconfinement phase, (3) ferromagnetic phase, (4) simple product phase. For $K_1/K_2\gg 1$, $K_1$-term is dominant. 
This condition gives Gauss' law constraint $\sigma^{x}_{j-1/2}X_{j}\sigma^{x}_{j+1/2}=1$ for the Hilbert space \cite{LGT_simu}. 
Under this condition, in the parameter region such as $J_1 > J_2$, $J_1$-cluster term is dominant, leading to the SPT phase protected by $Z_2 \times Z_2$ symmetry. 
This SPT phase is also interpreted as Higgs phase, where charges are condensed and a string order parameter (the open Wilson string) is finite as recently suggested in \cite{Verresen2022}.
While for $J_1 < J_2$, $J_2$-term is dominant, deconfinement phase of LGT emerges, which can be regarded as a  1D counterpart of the toric code in 2D. 
In this phase, two-fold degeneracy appears in the ground state by the long-range order $\langle\sigma^{x}_{j-1/2}\sigma^{x}_{j'+1/2}\rangle \neq 0$ via Gauss' law and finite magnetization (a finite charge density) $\langle X_j\rangle \neq 0$. 
Interestingly enough, this phase can be regarded as a spontaneously broken phase of the $W$-symmetry \cite{Verresen2022}.

Furthermore, $K_2 > K_1$ regime is also interesting, where Gauss' law is weakened, and other phases emerge.  
For $J_1 > J_2$, a ferromagnetic phase appears with spontaneous broken $Z_2$- symmetry since $\sigma^z_{j+1/2}$ is frozen and the model reduces to a transverse field Ising model~\cite{Borla2021}. 
While for $J_1< J_2$, the $J_2$ term is dominant and as a result, a trivial product state emerges, stabilized by $X_j$ and $\sigma^z_{j+1/2}$. 

With open boundary conditions, the above four ground states exhibit different characters \cite{Borla2021,Verresen2022}, in particular, the degeneracy of these ground states is different. 
In the Higgs=SPT phase, the ground state is four-fold degenerate due to the presence of a zero-energy edge mode at each edge. 
This is directly observed by counting the number of stabilizers stabilizing the state. 
For $K_1\to \infty$ and $J_1\to \infty$, the total number of the two stabilizers of $K_1$ and $J_1$ term is $2L-1$. 
This leads to two redundant degree of freedom, $L_t-(2L-1)=2$, inducing four-fold ($=2^{L_t-(2L-1)}$) degeneracy.
In the topological phase for $J_2/J_1 \gg 1$, the ground state is two-fold degenerate since the total number of matter site is smaller than that of the gauge link \cite{deconfinement_stab}.
For the ferromagnetic phase, the ground state is doubly degenerate since cat states occur.
For the product phase, the ground state is unique since the state is stabilized by all $K_2$ and $J_2$ terms. 
The above ground state degeneracy is one of the properties of the Hamiltonian. 
We shall show that states, which can be regarded as counterparts of the above four ground states, are produced by the MoC as mixed states by employing the mixed-state protocol.

It is expected that some of four phases can be characterized by bulk non-local order parameters \cite{Verresen2022}. 
For the Higgs=SPT phase, the bulk order can be characterized by the following decorated domain wall operator (DWO), 
\begin{eqnarray}
G(i_0,j_0)=Z_{i_0}\biggl(\prod^{j_0-1}_{j=i_0}\sigma^z_{j+\frac{1}{2}}\biggr)Z_{j_0},
\end{eqnarray} 
where $i_0$ and $j_0$ are two separated matter sites. 
The Higgs=SPT phase has a finite expectation value of $G(i_0,j_0)$. 
For the ferromagnetic phase, the bulk order can be characterized by spin-spin correlation operator, 
\begin{eqnarray}
S(i_0,j_0)=Z_{i_0}Z_{j_0},
\end{eqnarray} 
where $i_0$ and $j_0$ are two separated matter sites. 
The ferromagnetic phase has a finite expectation value of $S(i_0,j_0)$.
We shall apply modified version of these non-local order parameters to the numerics of the MoC as shown later on.

The global ground state phase diagrams were analytically studied and obtained in \cite{Borla2021} (See Fig.~8 in \cite{Borla2021}, where the phase diagrams of a related model to $H_{Z_2}$ are shown). 
There, the four phases are displayed in $(J_1/J_2)$-$(K_1/K_2)$ plane, and two phase boundaries are given by $J_1/J_2=1$ and $K_1/K_2=1$.

\begin{figure}[t]
\begin{center} 
\includegraphics[width=8.5cm]{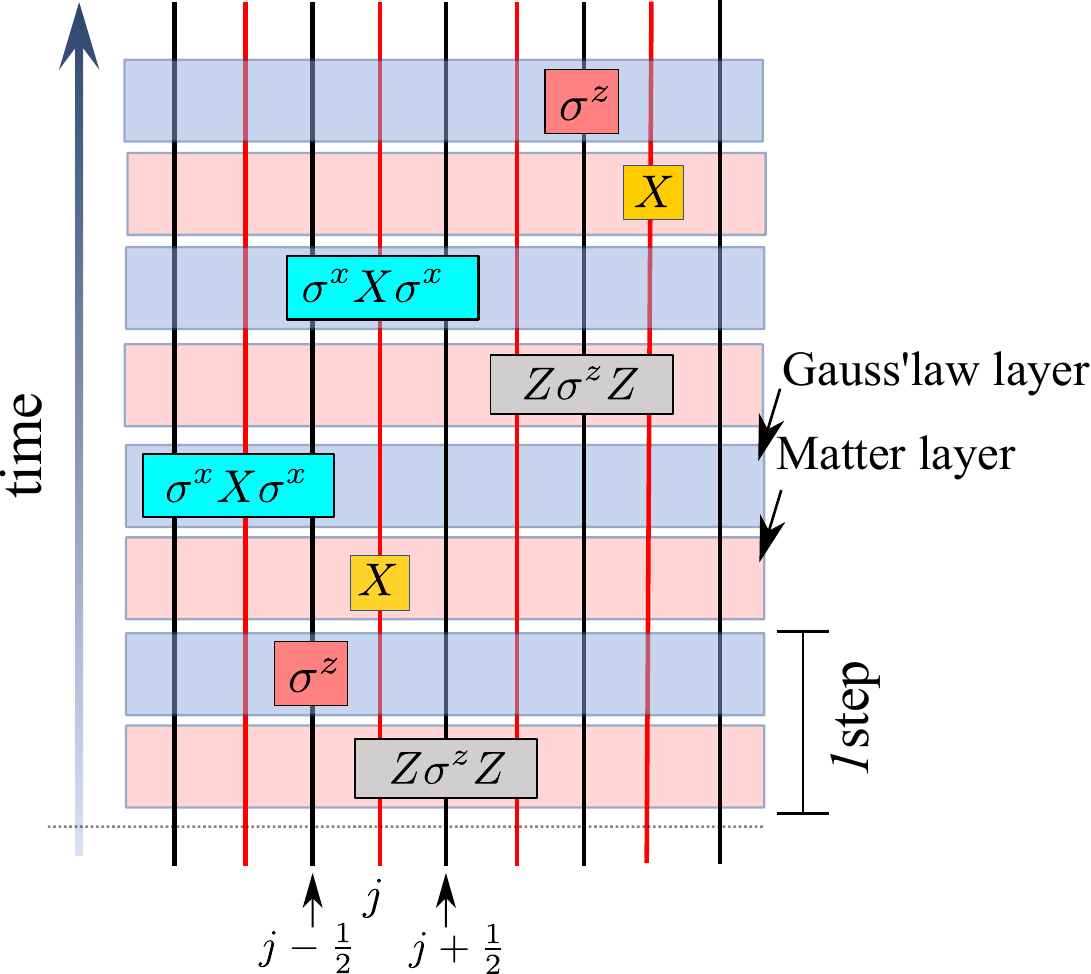}  
\end{center} 
\caption{Schematic figure of the measurement-only circuit corresponding to the Hamiltonian $H_{Z_2}$. The black and red lines represent the matter sites and gauge links, respectively. 
The blue and magenta shaded layers represent Gauss' law and matter layers, respectively. 
The one time step includes the two layers.
}
\label{Fig3}
\end{figure}

\subsection{Circuit setup corresponding to $H_{Z_2}$}
We setup a MoC, which is expected to produce a qualitatively the same phase diagram with that of the ground state reviewed in the previous subsection. 
To this end, we introduce a two-layered projective measurement in a single time step as shown in Fig.~\ref{Fig3}. The two layers are composed of matter layer and Gauss' law layer.

We consider system of $L$ matter sites with open boundary conditions, where $L+1$ gauge links exist as in Fig.~\ref{Fig2}. 
The total degree of freedom is therefore $L_t$.
Here, we introduce four different types of stabilizers and corresponding projective measurements, 
which are defined as 
\begin{eqnarray}
{\hat M}^{1a}_{j}&=&\sigma^{x}_{j-1/2}X_{j}\sigma^{x}_{j+1/2},\:\:\: 
{\hat M}^{1b}_{j'}=\sigma^z_{j'+1/2},\\
{\hat M}^{2a}_{j}&=&Z_j\sigma^{z}_{j+1/2}Z_{j+1},\:\:\: 
{\hat M}^{2b}_{j}=X_{j},
\end{eqnarray}
where $j=0,1,\cdots, L-1$ and $j'=-1,0,\cdots, L-1$.
The above four kinds of operators are included in $H_{Z_2}$ and they satisfy properties of stabilizer, i.e., 
$[\hat{M}^{k \alpha}_{i},\hat{M}^{k \alpha}_{j}]=0$ and  $(\hat{M}^{k \alpha}_j)^2=1$ for $k=1,2$ and $\alpha=a,b$, and note that $\hat{M}^{k a}_{i}$ and $\hat{M}^{k b}_{j}$  anti-commute with each other for a pair of $(i,j)$. That is, projective measurements of $\hat{M}^{k a}_{i}$ and $\hat{M}^{k b}_{j}$ for $k=1,2$ are competitive with each other. 
In the MoC, for each matter layer, we apply stabilizers ${\hat M}^{2a}_{j}$ and ${\hat M}^{2b}_{j}$ with probability $p^A$ and $p^B$, respectively, with $p^A+p^B=1$. 
The measured site $j$ is chosen randomly with equal probability, similar to the case in Sec.II.B. 
In each Gauss' law layer, we apply stabilizers ${\hat M}^{1a}_{j}$ and ${\hat M}^{1b}_{j}$ with probability $p^C$ and $p^D$, respectively, with $p^C+p^D=1$. 
The measured site $j$ is chosen again randomly with equal probability. 

Since each ground state of $H_{Z_2}$ is degenerate in a different manner, the mixed-state update procedure is efficient to detect and characterize the phases since we can count the number of stabilizers directly in numerics. 
In the previous subsection, we expect that the PRC holds even for the mixed-state protocol. 
For the practical MoC, we set the infinite-temperature state as an initial state, and then the mixed state is evolved for a large number of discrete time steps. 
Another reason to employ the mixed-state protocol is that the initial-state dependence existing in the pure state update can be avoided. 

We consider a long time evolution with the total number of steps $t_N=4(2L+1)$. 
In general, the initial mixed state is purified by projective measurements. 
We first obtain a steady state (mixed or pure state) in each  measurement pattern (a single stochastic process) and calculate physical observables in the steady state. 
Then, we gather many samples of steady states and physical observables as an ensemble and investigate the properties of the ensemble to compare them with the ground state properties of the target Hamiltonian.

\section{Numerical Results of purification dynamics}
In this section, we shall show numerical demonstrations of the MoC defined in the previous section, and verify that the MoC  generates steady stabilizer states, the phase diagram of which is similar to the ground state phase diagram of $H_{Z_2}$. 

\subsection{Explanation of numerical calculation}
We make use of stabilizer update numerical algorithm \cite{Gottesman1997,Aaronson2004,Nielsen_Chuang} to simulate the MoC. 
In particular, we employ the mixed-state update methods of stabilizer dynamics employed in \cite{Gullans2020,Ippoliti2021}, in which information of sign for updating stabilizers is not stored.

We start with the state at infinite temperature $\rho=\frac{1}{N_D}\hat{I}$, where $N_D$ is the Hilbert space dimension of the system ($N_D=2^{L_t}$). 
Generally, the time evolution by sequential projective measurements of stabilizers makes the initial mixed state purified (the rank of the density matrix is decreasing.) 
For a long time period, a purified state emerges as a steady state, but it cannot be necessarily a genuine pure state, i.e., it is allowed to be a mixed state. 
We expect that a steady mixed state corresponds to a multiplet of the ground states of $H_{Z_2}$. 
More precisely, the rank of a steady mixed state denoted by $N_{cs}$ is related to the degeneracy of the ground state of $H_{Z_2}$ denoted by $N_{gd}$, as $2^{N_{cs}}=N_{gd}$. 
In the stabilizer formalism, the rank is related to the dimension of code space $2^{N_{cs}}$ with $N_{cs} \equiv L_t-N_R$, where $N_R$ is total number of linearly-independent stabilizers generating the mixed state \cite{QI_text}.

\begin{figure}[t]
\begin{center} 
\includegraphics[width=7.5cm]{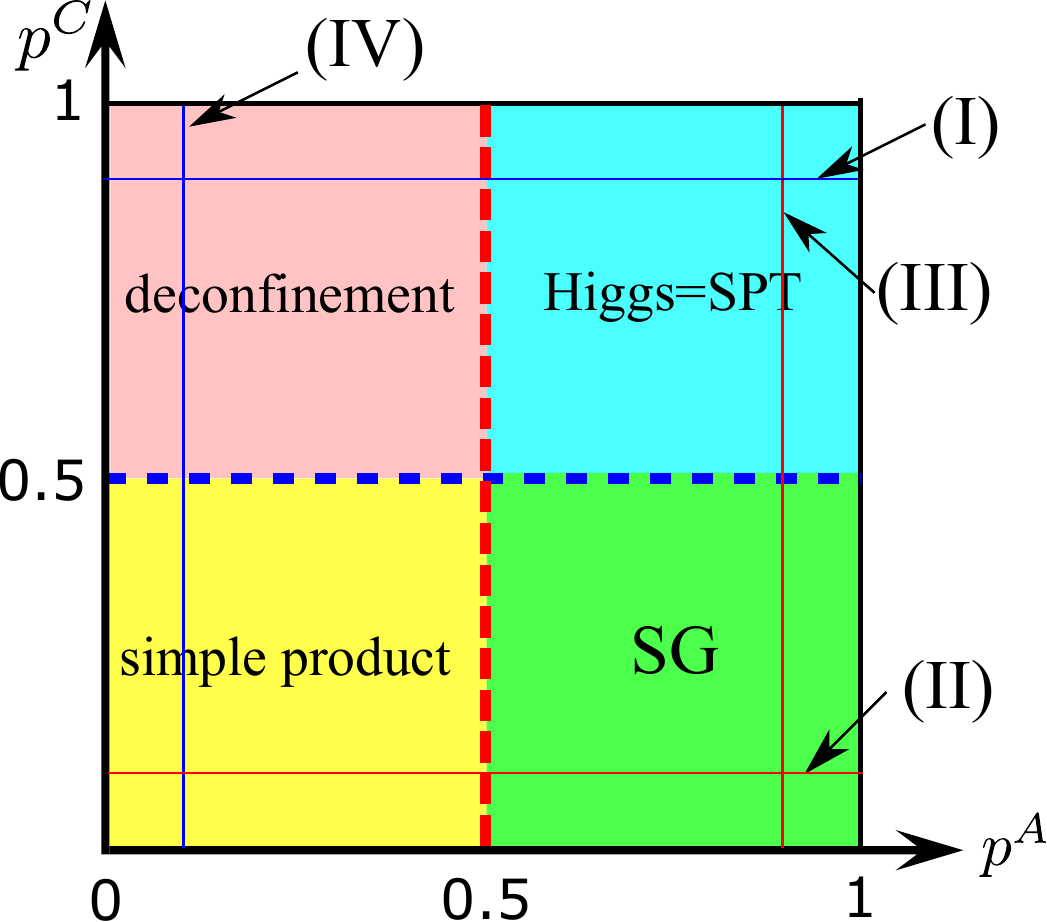}  
\end{center} 
\caption{Schematic figure of phase diagram obtained by the MoC. 
Mixed-state algorithm is employed. 
The red and blue dashed lines are obtained phase boundaries, $p^A\approx 0.5$ and $p^C\approx 0.5$
in the present work.
The red and blue solid lines represent 
the typical parameter sweeps studied in detail. 
}
\label{Fig4}
\end{figure}

\begin{figure*}[t]
\begin{center} 
\includegraphics[width=18cm]{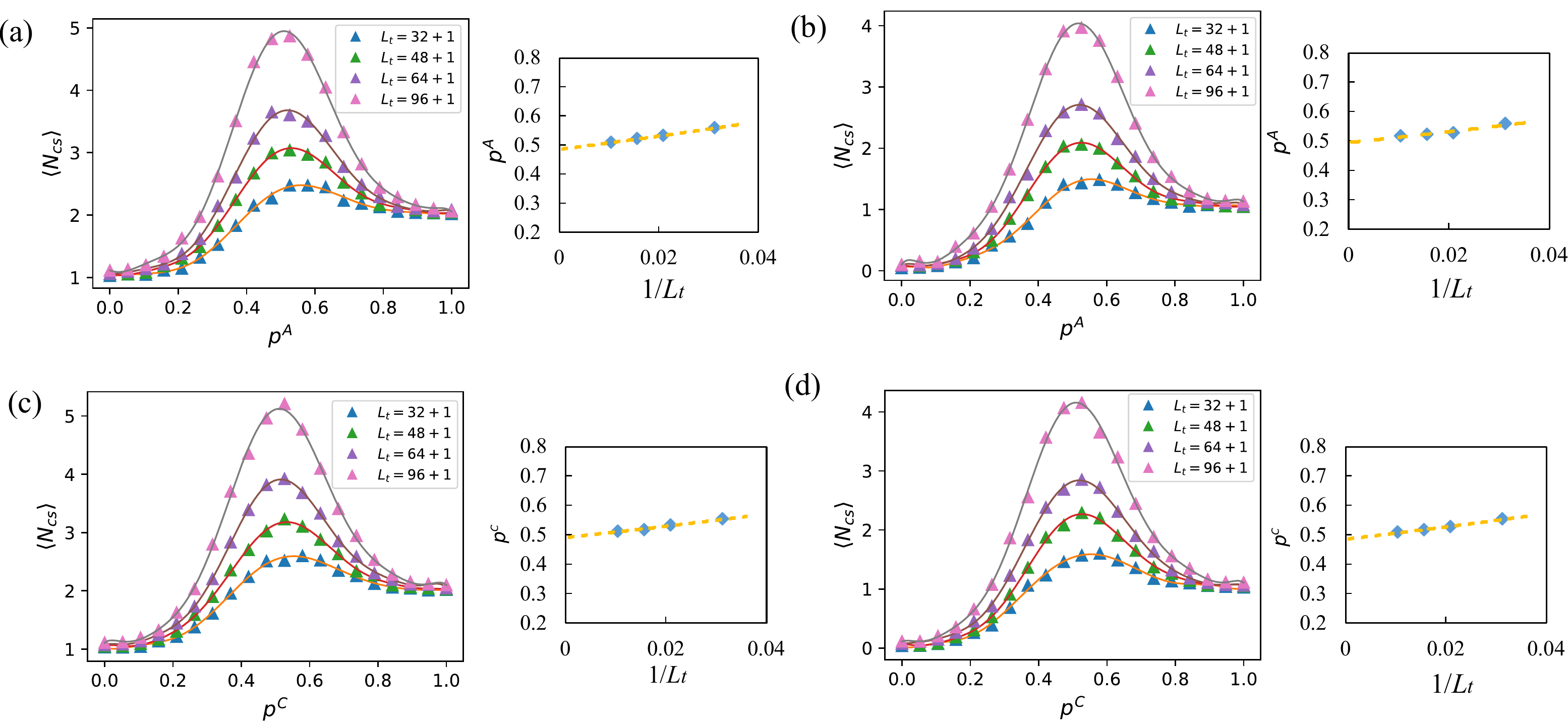}  
\end{center} 
\caption{Phase transition behaviors of $\langle N_{cs}\rangle$ for various system sizes. The solid lines are fitting lines. 
(a) The data for a strong Gauss' law case, $p^C=0.9$.
(b) The data for a weak Gauss' law case, $p^C=0.1(p^D=0.9)$, where the gauge dynamics is frozen, $\sigma^{z}_{j+1/2}\to 1$.
(c) The data for a Gauss' law sweep case with a fixed $p^A=0.9$.
(d) The data for a Gauss' law sweep case with a fixed $p^A=0.1$.
Right small panels: System-size dependence of $p^A_c$ and $p^C_c$ deduced from the peak of the fitting lines. 
We can extrapolate the critical probabilities for $1/L\to 0$. Here we used exponential fitting line.}
\label{Fig5}
\end{figure*}

In practical calculation of  the target observables shown later, we employ $400-600$ different measurement patterns for various system sizes and various values of probabilities, and take an ensemble average of saturation values of the observables at $t_N=4L$, where the state reaches a steady state (mixed or pure state).

\subsection{Physical observables}
To identify phase of the state obtained by the MoC, 
we first observe the degree of the code space $N_{cs}$ obtained by counting the total number of linearly-independent stabilizers $N_{R}$. 
In particular for steady states, we calculate the ensemble average of it, denoted by $\langle N_{cs}\rangle$, obtained through many samples of the measurement patterns. 
In fact, $\langle N_{cs}\rangle$ is related to the average entropy of the state \cite{Gullans2020}  and also is expected to relate to the degeneracy of the ground state of the corresponding Hamiltonian $H_{Z_2}$, as we explained in the above. 

Furthermore, to examine if the MoC dynamics generates the Higgs=SPT or ferromagnetic phase in the bulk as a steady state, we calculate decorated domain-wall order (DWO)~\cite{Lavasani2021,Verresen2022}, which is defined as follows, 
\begin{eqnarray}
({\rm DWO})^2 \equiv 2^{N_{cs}}\mathrm{tr}[\rho(t_N)G(i_0,j_0)\rho(t_{N})G(i_0,j_0)].
\end{eqnarray}
where 
$$
G(i_0,j_0)=Z_{i_0}\biggl(\prod^{j_0-1}_{j=i_0}\sigma^z_{j+\frac{1}{2}}\biggr)Z_{j_0}.
$$
Here, by using linearly-independent stabilizer generators, the density matrix of the system state (mixed state) is given by 
\begin{eqnarray}
\rho(t_N)=\prod^{N_R-1}_{\ell=0}\biggr( \frac{1+s^{\ell}(t_N)}{2}\biggl),
\end{eqnarray}
where $s^{\ell}(t_N)$ is $N_R$'s updated stabilizers (linearly-independent). 
In the LGT, $G(i_0,j_0)$ is nothing but a gauge-invariant correlator of matter field (Higgs field) connected by  Wilson string.
On the other hand, $({\rm DWO})^2$ is a kind of Edward-Anderson type string order to detect SPT phase \cite{Lavasani2021}.

We further calculate the following spin-glass long-range order parameter (SGO) to characterize ferromagnetic phase,
\begin{eqnarray}
({\rm SG})^2 \equiv 2^{N_{cs}}\mathrm{tr}[\rho(t_N)S(i_0,j_0)\rho(t_N)S(i_0,j_0)],
\label{SGO}
\end{eqnarray}
where $S(i_0,j_0)=Z_{i_0}Z_{j_0}$.
In the update of the MoC without storing information of sign of the stabilizers, the ferromagnetic phase implies the presence of spin-glass like phase, thus, in what follows, we call the phase ``spin-glass (SG) phase" instead of ferromagnetic phase.
The further practical calculation scheme in our numerics is explained in Appendix B.

\subsection{Phase diagram of steady state obtained by the MoC}
We start with observing $\langle N_{cs}\rangle$. 
From the behavior of $\langle N_{cs}\rangle$, we verify that the MoC generates four different kinds of steady states and find the qualitative phase diagram in $p^A$-$p^C$ plane as shown in Fig.\ref{Fig4}. 
This phase diagram is very close to the ground state phase diagram of the Hamiltonian $H_{Z_2}$ proposed in \cite{Borla2021,Verresen2022}, in which two phase boundaries exist at $J_1/J_2=1$ and $K_1/K_2=1$ separating the ground state of the system $H_{Z_2}$. 
The phase diagram of the steady state obtained by our numerics of the MoC has also two phase boundaries at $p^{A}/p^{B}\sim 1$ and $p^{C}/p^{D}\sim 1$. 
Therefore, our study confirms the PRC, i.e., the MoC with a suitable setting of projective measurement of stabilizers can generate (mixed) steady states that are very close to the gauge-theoretical ground states of $H_{Z_2}$ through long but finite-period evolution by the MoC.

We investigate details of the transition properties of mixed states in the MoC. 
The behavior of $\langle N_{cs}\rangle$ along the four typical lines in the parameter space (I)-(IV), displayed in Fig.~\ref{Fig4}, is observed. 
The results for various system sizes are shown in Fig.~\ref{Fig5}. 
We find that all data exhibit clear system-size dependence and the peaks of $\langle N_{cs}\rangle$ are located in the vicinity of $p_A$ or $p_c\sim 0.5$. 
These peaks are obviously a signature of the phase transition. 

Calculations in Fig.~\ref{Fig5} (a) are for the case of  $p^C=0.9$, in which Gauss' law is enforced strongly. 
We find the value of $\langle N_{cs}\rangle$ clearly changes $1\to 2$ as increasing $p^{A}$. 
This indicates that the mixed state exhibits transition from the deconfinement phase to the Higgs=SPT phase since $\langle N_{cs}\rangle=2$ shows the presence of four-fold degenerate stabilizer states, corresponding to the ground state degeneracy of the Higgs=SPT phase of $H_{Z_2}$ in open boundary case, whereas $\langle N_{cs}\rangle=1$ corresponds to two-fold degenerate states by the spontaneous breaking of the magnetic symmetry in $H_{Z_2}$.  
See the data of Fig.~\ref{Fig5} (b) for $p^C=0.1(p^D=0.9)$. 
Gauss' law is weak and the gauge variable is frozen as $\sigma^z_{j+1/2}\to 1$ instead. 
We observe that the value of $\langle N_{cs}\rangle$ clearly changes $0\to 1$ as increasing $p^{A}$. 
This implies that the state changes from the product pure state stabilized by all $X_j$ and $\sigma^z_{j+1/2}$ to the SG phase, which are two-fold degenerate states stabilized by $Z_j\sigma^{z}_{j+1/2}Z_{j+1} \to Z_{j}Z_{j+1}$, corresponding to the ground state degeneracy of the ferromagnetic phase (cat states) of $H_{Z_2}$. 
Next, see the data of Fig.~\ref{Fig5} (c) where we fix $p^A=0.9$ and vary the strength of Gauss' law. 
We observe that the value of $\langle N_{cs}\rangle$ clearly changes $1\to 2$ as increasing $p^{C}$ implying that the SG phase transitions into the Higgs=SPT phase. 
Also, see the data of Fig.~\ref{Fig5} (d) where we fix $p^A=0.1$ and vary the strength of Gauss' law. 
We observe that the value of $\langle N_{cs}\rangle$ clearly changes $0\to 1$ as increasing $p^{C}$ implying the simple product pure state transitions into the deconfinement phase.

We further analyze the system-size dependence of $\langle N_{cs}\rangle$ along the above mentioned four lines in the parameter space. 
From the data, we can obtain phase transition points in the MoC. 
We fit the data points of $\langle N_{cs}\rangle$ \cite{Ncs_fit} and deduce the location of the peak of the fitting line with the corresponding probability for each system size. 
The exponential fitting \cite{exp_fit} of selected probability points of different system sizes is performed on $1/L$-axis, and the fitting line is extrapolated to estimate the transition probability point $p^{A}_c$ or $p^C_{c}$ for $L_t\to \infty$. 
These FSS data are displayed in the right panels in (a)-(d) of Fig.~\ref{Fig5}. 
By using this method, we estimate the phase transition points: for the line (I), $p^A_c=0.485(8)$, the line (II), $p^A_c=0.494(7)$, the line (III), $p^C_c=0.491(1)$, and the line (IV), $p^C_c=0.485(6)$. 
The above values are fairly close to $0.5$, implying that $p^A/p^B\sim 1$ and $p^C/p^D\sim 1$ are phase boundaries in the MoC. 
The results of estimation indicate the validity of the PRC for the phase boundaries, $p^A/p^B \longleftrightarrow J^{1}/J^2$ and $p^C/p^D \longleftrightarrow K^{1}/K^2$. 
Note that these estimated values are slightly smaller than $0.5$, and we expect that the reason for that comes from the difference of the total number of the stabilizers corresponding to each phase. 
In Appendix C, we further show the behavior of $\langle N_{cs}\rangle$ on other lines in the parameter space and determine a quadruple critical transition point.

\begin{figure}[t]
\begin{center} 
\includegraphics[width=9cm]{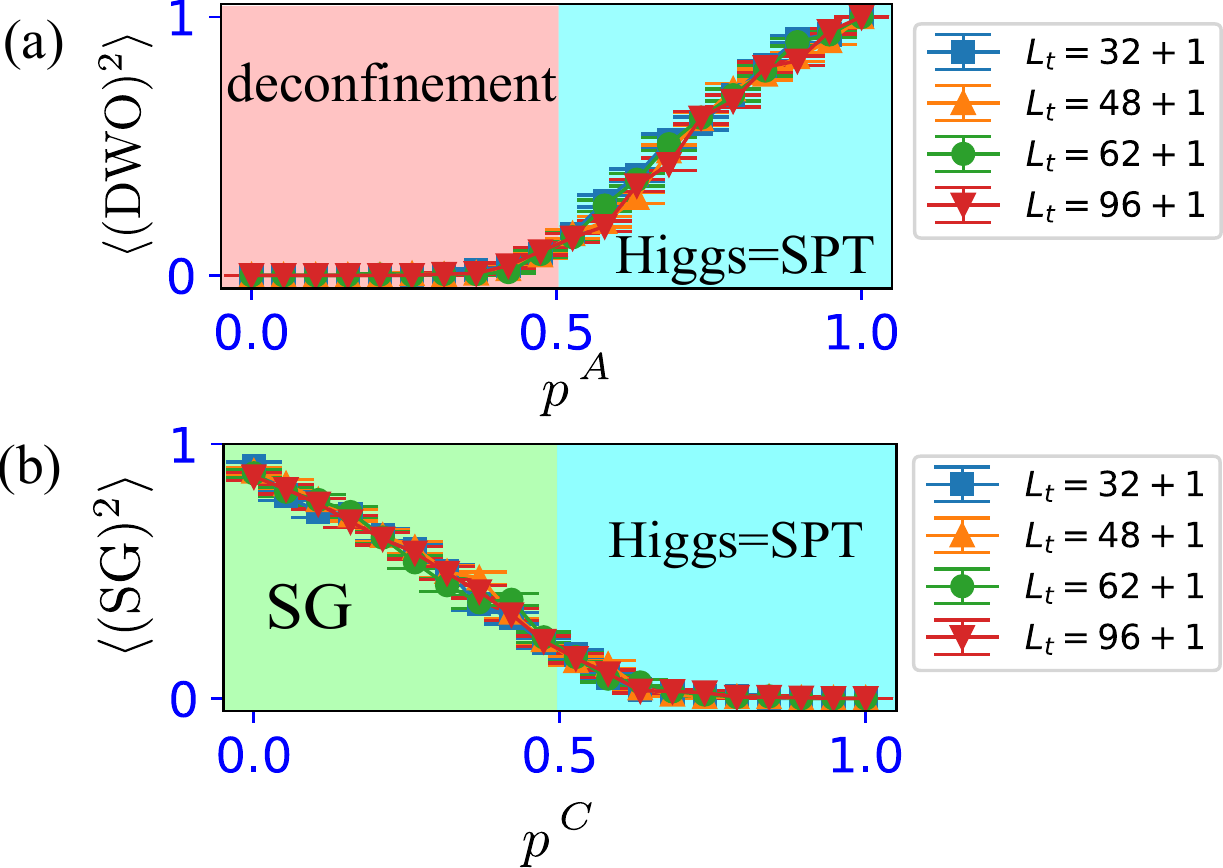}  
\end{center} 
\caption{(a) Decorated domain wall operator (DWO) with fixed value $p^C=0.9$ (Strong Gauss' law is enforced). Decorated domain wall is condensed in the Higgs=SPT phase indicating SSB of matter parity symmetry. (b) Spin-glass order (SGO) with fixed value $p^A=0.9$. 
Both data indicate no system size dependence.}
\label{Fig6}
\end{figure}

In addition, we calculate the average values of the DWO and SGO along the lines (I) and (II) in Fig.~\ref{Fig4}, where we set $i_0=1$ and $j_0=L-2$. 
The results are shown in Fig.~\ref{Fig6} (a) and \ref{Fig6} (b). 
The DWO becomes finite in the Higgs=SPT regime ($p^A\gtrapprox 0.5$) and the SGO becomes finite in the SG ferromagnetic regime ($p^C\lessapprox 0.5$). 
These results support the existence of the bulk Higgs=SPT and SG phases produced by the MoC. 
We also find that the behaviors of the DWO and SGO are insensitive to the system size.

Finally, we estimate the criticality of the transitions in the MoC simulation. 
Before showing the numerical results, it should be remarked that the present simulation of mixed-state update in the MoC has aspects different from the usual ground state simulation governed by the Hamiltonian. 
That is, the critical exponents of the mixed-state transition in the MoC might be different from those of the genuine ground state phase transition emerging by varying the parameters in Hamiltonian. 
The criticality of the ground state phase transitions in the system $H_{Z_2}$ was investigated in \cite{Borla2021,Verresen2022} in terms of conformal field theory (CFT). 
However, our finding phase transition is not necessarily governed by such a CFT. 
At present, it is not clear if the pure-state and mixed-state updates have the same criticality, even though the transition points are the same.
This is an interesting future problem. 
Keeping this remark in mind, 
we carry out FSS analysis for the (I) and (III) lines in Fig.~\ref{Fig4}, that is, we consider the deconfinement-Higgs=SPT phase and the SG phase-Higgs=SPT phase transitions. 

To estimate its criticality (critical exponents), we apply the FSS analysis to $\langle N_{cs}\rangle$. 
Here, we employ the following scaling ansatz \cite{Szyniszewski2019,Takashima2005},
\begin{eqnarray}
\langle N_{cs}\rangle (p^\alpha, L)=L^{\frac{\gamma}{\nu}}\Psi((p^\alpha-p^{\alpha}_c)L^{1/\nu}),
\end{eqnarray}
where $\Psi$ is a scaling function, $\gamma$ and $\nu$ are critical exponents and $p^{\alpha=A,C}_c$ is a critical transition probability. 
We use the extrapolated values of $p^{A(C)}_{c}$ for $L_t\to \infty$ shown in Fig.~\ref{Fig5} (a) and ~\ref{Fig5} (c), and determine the scaling function $\Psi$ by searching the optimal values of $\gamma$ and $\nu$. 
There, by using the data of $\langle N_{cs}\rangle$, the fitting curve for the scaling function is obtained via a 12-th order polynomial function with the optimal coefficients for various values of $\gamma$ and $\nu$, and then the coefficient of determination $R^2$ is estimated to find optimal $\gamma$ and $\nu$. 

\begin{figure}[t]
\begin{center} 
\includegraphics[width=7.5cm]{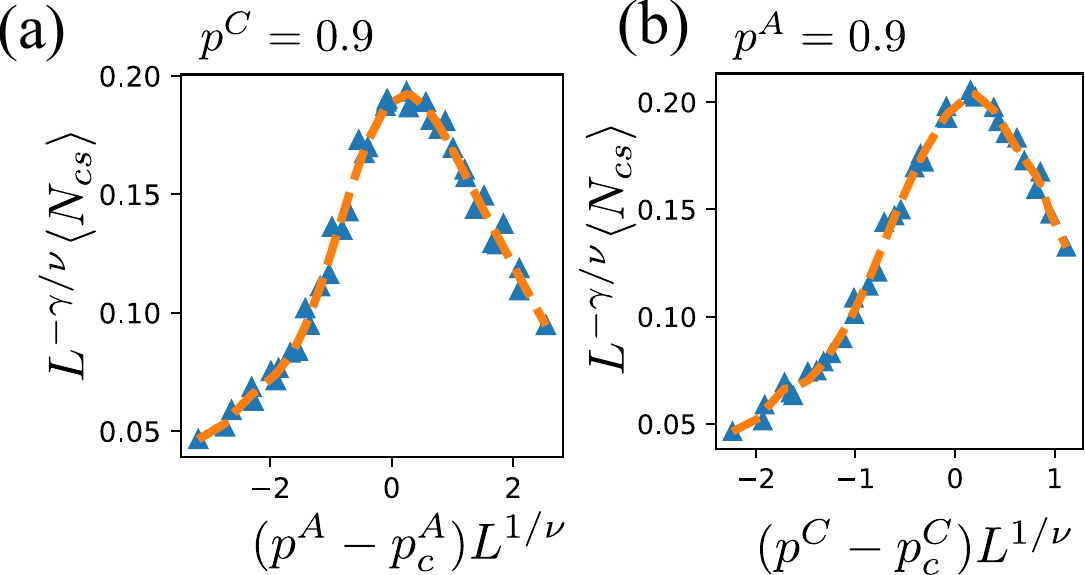}  
\end{center} 
\caption{Best optimal scaling functions for the transition behavior of the sweeps (I) and (III) in Fig.~\ref{Fig4}. 
The curves are obtained from the optical fitting calculation by using the estimated values, $p^A_c=0.485(8)$ and $p^C_c=0.491(1)$ from the data in Fig.~\ref{Fig5} (a) and ~\ref{Fig5} (c). 
In the data (a), the optimal critical exponents are $\gamma=1.53(0)$ and $\nu=2.15(0)$. 
The fitting line of the scaling function has $R^2=0.990(4)$. 
In the data (b), the optimal critical exponents are $\gamma=1.85(0)$ and $\nu=2.6(0)$. 
The fitting line of the scaling function has $R^2=0.995(8)$. 
For both data, the fitting curves of the scaling function is obtained via a 12-th order polynomial function with the best optimal coefficients.}
\label{Fig7}
\end{figure}

The scaling functions obtained by this FSS analysis are displayed in Figs.~\ref{Fig7}(a) and \ref{Fig7} (b) where we used $L=48, 64, 96$ data points in Fig.~\ref{Fig5} (a) and \ref{Fig5} (c) and set $p^A_c=0.485(8)$ and $p^C_c=0.491(1)$ for the parameter sweeps (I) and (III), respectively. 

For the deconfinement-Higgs=SPT phase transition, the optimal critical exponents are estimated as $\gamma=1.53(0)$ and $\nu=2.15(0)$. 
The fitting line of the scaling function has $R^2=0.990(4)$. 
For the SG phase-Higgs=SPT phase transition, the optimal critical exponents are estimated as $\gamma=1.85(0)$ and $\nu=2.6(0)$. 
The fitting line of the scaling function has $R^2=0.995(8)$. 
We should not compare these values with those of CFTs since our target phase transition is for mixed states and occurs in the MoC, as we explained in the above. 
Furthermore, the criticality observed in the present MoC for the mixed states may reveal some non-trivial aspects of the symmetry enriched topological phase transition. 
Anyway, to clarify physical meanings of the obtained critical exponents is a future problem.

We summarize the results of our numerical calculation of the MoC, where we used mixed-state update methods efficient to study the various degenerate ground-state mutiplets. 
We numerically demonstrated that the PRC between the gauge-theory Hamiltonian $H_{Z_2}$ and the corresponding MoC holds in the phase diagram level (that is, phase boundaries) even though mixed-state update simulation is employed, but the criticality is different. 
Conversely, based on the PRC guiding principle, the MoC with a suitable set of stabilizer projective measurements can produce various stabilizer states corresponding to the interesting ground states of the gauge theory. 

\section{Conclusion}
\uline{In the first half of this work}, 
we focused on a Hamiltonian including different types of stabilizers of Eq.~(\ref{stab_H}) and gave a qualitative argument of the PRC by comparing the propagators obtained from the imaginary-time path integral and the ensemble average of the MoC. 
In particular, we showed two concrete examples supporting and strengthen the validity of the PRC. We also discussed that the PRC can be extended to the mixed-state dynamics since the PRC is based on the structure of the propagators, itself. 
Needless to say, the discussion on the PRC in this work is qualitative, and more rigorous and mathematical proof for this conjecture is an important future problem and welcome.

\uline{In the second half of this work}, to examine the validity and utility of the PRC, we investigated the (1+1)-D $Z_2$ lattice gauge-Higgs model, which includes very rich physics and distinct degenerate ground-state multiplet for each phase, by the practical use of the MoC. 
We showed that the MoC with suitable stabilizer projective measurements and suitable probability ratios produces a steady-state phase diagram, which is quite similar to the ground state phase diagram of the  corresponding gauge-Higgs Hamiltonian previously studied in \cite{Borla2021,Verresen2022}.
Our numerical result of the MoC is a concrete example indicating that (I) the PRC is observed as far as the phase structure even in the under mixed-state update, extending and corroborating the analytical conjecture of the pure state update, 
(II) the PRC can be a good guiding principle to produce interesting and desired states (including mixed state) by MoCs with suitable stabilizer projective measurement suggested by the PRC. 
As a specific concrete example, our MoC demonstrate the presence of Higgs=SPT phase and other symmetry-breaking type orders such as SG phase by controlling the strength of Gauss' law with varying the measurement probability.

Finally, even though this work mainly studied gauge theory in $(1+1)$D as a concrete example, it is straightforward to apply the present methods to other quantum systems in higher dimensions. 
We hope that we will report studies on them in a future.

\section*{Acknowledgements}
This work is supported by JSPS KAKEN-HI Grant Number JP21K13849 (Y.K.). 

\appendix
\widetext
\section*{Appendix A: Three spin cluster model}
We consider the following Hamiltonian,  
$$
H'_{\rm stab2}=\sum^{2}_{j=0}[-J_1Z_{j-1}X_jZ_{j+1}-J_2X_j],
$$
where periodic boundary conditions are imposed.

We focus on the imaginary-time propagator $G_{j+1,j}=\langle\ell|e^{-\delta\tau H}|\ell'\rangle$. 
Here, we consider a set of basis based on $X_j$ 
\begin{eqnarray}
\{|\ell\rangle\}&=&\{|+++\rangle, \:\:|-++\rangle, \:\:
|+-+\rangle, \:\:|++-\rangle,\:\:\nonumber\\
&&|--+\rangle, \:\:|-+-\rangle,\:\:
|+--\rangle, \:\:|---\rangle
\}.
\end{eqnarray}
Then with this basis, the full matrix form of $G_{j+1,j}$ is obtained as
\begin{eqnarray}
&&(G_{j+1,j})\approx \nonumber\\
&&\begin{bmatrix}
    c^1[a^3-b^3]&             0&             0&             0&   c^3[ab(a-b)]&  c^3[ab(a-b)]&  c^3[ab(a-b)]&             0\\
               0&  c^2[a^3+b^3]&  c^2[ab(a+b)]&  c^2[ab(a+b)]&              0&             0&             0& c^4[-ab(a+b)]\\
               0&  c^2[ab(a+b)]&  c^2[a^3+b^3]&  c^2[ab(a+b)]&              0&             0&             0& c^4[-ab(a+b)]\\
               0&  c^2[ab(a+b)]&  c^2[ab(a+b)]&  c^2[a^3+b^3]&              0&             0&             0& c^4[-ab(a+b)]\\
    c^1[ab(a-b)]&             0&             0&             0&   c^3[a^3-b^3]& c^3[-ab(a-b)]& c^3[-ab(a-b)]&             0\\
    c^1[ab(a-b)]&             0&             0&             0&  c^3[-ab(a-b)]&  c^3[a^3-b^3]& c^3[-ab(a-b)]&             0\\
    c^1[ab(a-b)]&             0&             0&             0&  c^3[-ab(a-b)]& c^3[-ab(a-b)]&  c^3[a^3-b^3]&             0\\
               0& c^2[-ab(a+b)]& c^2[-ab(a+b)]& c^2[-ab(a+b)]&              0&             0&             0&  c^4[a^3+b^3]\\
\end{bmatrix},\nonumber\\
\end{eqnarray}
where
\begin{eqnarray}
&&a=\cosh(\delta\tau J_1),\:\: b=\sinh(\delta \tau J_1),\nonumber\\
&&c^1=e^{3\delta \tau J_2},\:\:c^2=e^{\delta \tau J_2},\:\:c^3=e^{-\delta \tau J_2},\:\:c^4=e^{-3\delta \tau J_2}.\nonumber
\end{eqnarray}

We turn to the propagator of the MoC. 
By employing the set of basis $\{|\ell\rangle\}$ above, the full matrix form of $\overline{I_{m+1,m}}$ is obtained as 
\begin{eqnarray}
(\overline{I_{m+1,m}})=
\begin{bmatrix}
    p^A/2+p^B&             0&             0&             0&        p^A/6&        p^A/6&         p^A/6&      0\\
            0&  p^A/2+2p^B/3&         p^A/6&         p^A/6&            0&            0&             0& -p^A/6\\
            0&         p^A/6&  p^A/2+2p^B/3&         p^A/6&            0&            0&             0& -p^A/6\\
            0&         p^A/6&         p^A/6&  p^A/2+2p^B/3&            0&            0&             0& -p^A/6\\
        p^A/6&             0&             0&             0& p^A/2+p^B/3&       -p^A/6&        -p^A/6&      0\\
        p^A/6&             0&             0&             0&       -p^A/6& p^A/2+p^B/3&        -p^A/6&      0\\
        p^A/6&             0&             0&             0&       -p^A/6&       -p^A/6&  p^A/2+p^B/3&      0\\
            0&        -p^A/6&        -p^A/6&        -p^A/6&            0&            0&             0&  p^A/2\\
\end{bmatrix}.\nonumber\\
\end{eqnarray}
We compare the components of the two matrices $(G_{j+1,j})$ and $(\overline{I_{m+1,m}})$ to obtain the following relationships:
\begin{eqnarray}
\mbox{1 column:}\:&& c^1 [a^3-b^3] \longleftrightarrow p^A/2+p^B, \:\: c^1[ab(a-b)]\longleftrightarrow p^A/6,\nonumber\\
\mbox{2-4 column:}\:&& c^2[a^3+b^3] \longleftrightarrow p^A/2+2p^B/3, \:\: c^2[ab(a+b)]\longleftrightarrow p^A/6,\nonumber\\
\mbox{5-7 column:}\:&& c^3[a^3-b^3] \longleftrightarrow p^A/2+p^B/3, \:\: c^3[ab(a-b)]\longleftrightarrow p^A/6,\nonumber\\
\mbox{8 column:}\:&& c^4[a^3+b^3] \longleftrightarrow p^A/2, \:\: c^4[ab(a+b)]\longleftrightarrow p^A/6.\nonumber
\end{eqnarray}

We proceed further approximation. Since $\delta \tau \ll 1$, we expand each component up to the order $\mathcal{O}(\delta \tau)$, then we find that the above eight relations reduce to
\begin{eqnarray}
\mbox{1 column:}\:&& 1+3\delta \tau J_2 \longleftrightarrow p^A/2+p^B, \:\: \delta\tau J_1\longleftrightarrow p^A/6,\nonumber\\
\mbox{2-4 column:}\:&& 1+\delta \tau J_2 \longleftrightarrow p^A/2+2p^B/3, \:\: \delta\tau J_1\longleftrightarrow p^A/6,\nonumber\\
\mbox{5-7 column:}\:&& 1-\delta \tau J_2 \longleftrightarrow p^A/2+p^B/3, \:\: \delta\tau J_1\longleftrightarrow p^A/6,\nonumber\\
\mbox{8 column:}\:&& 1-3\delta \tau J_2 \longleftrightarrow p^A/2, \:\: \delta\tau J_1\longleftrightarrow p^A/6.\nonumber
\end{eqnarray}
By requiring $G_{j+1,j}=C_0(\overline{I_{m+1,m}})$,  $\delta\tau J_1=\frac{C_0}{6}p^A$, and the relations $1+3\delta \tau J_2 = C_0(p^A/2+p^B)$ and $1+\delta \tau J_2 = C_0(p^A/2+2p^B/3)$ leads $\delta \tau J_2=\frac{C_0}{6}p^B$.
We obtain
\begin{eqnarray}
\frac{J_1}{J_2} \longleftrightarrow \frac{p^A}{p^B}.
\end{eqnarray}
\begin{figure}[b]
\begin{center} 
\includegraphics[width=6.5cm]{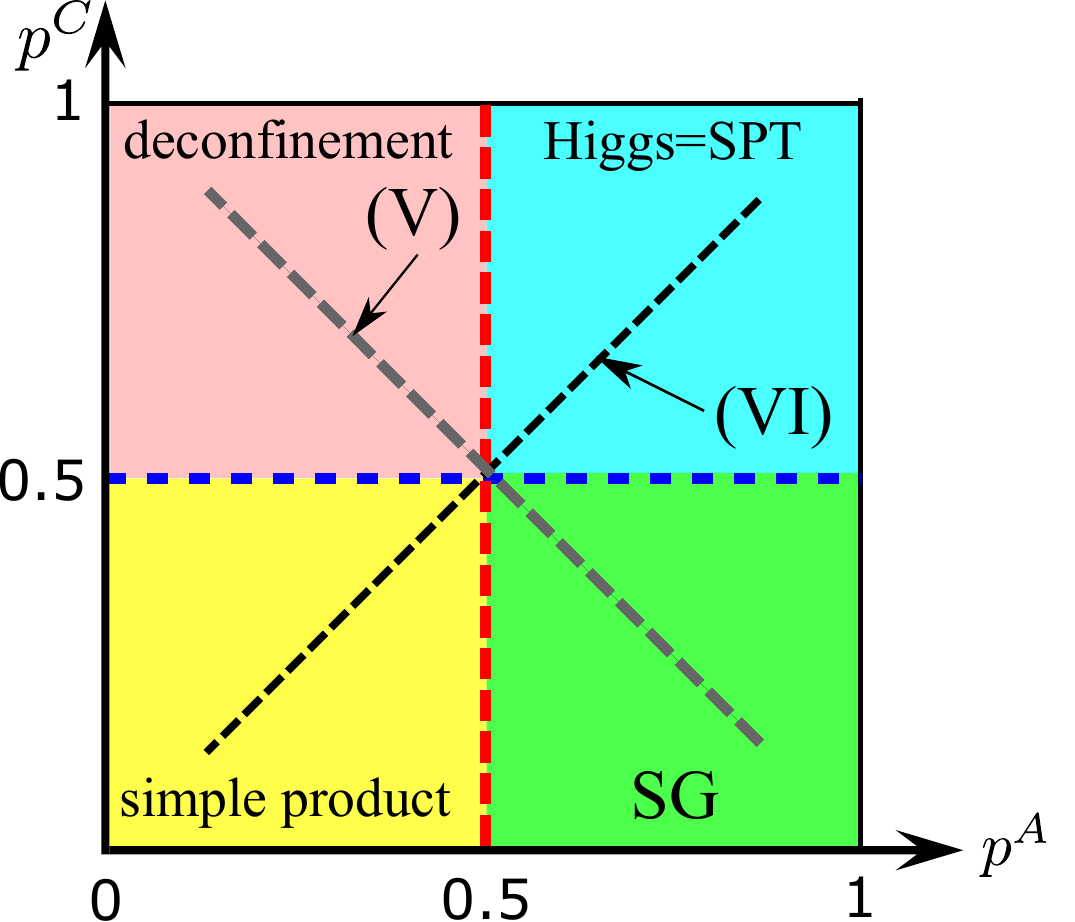}  
\end{center} 
\caption{Schematic figure of phase diagram obtained by the MoC.
Mixed-state algorithm is employed. The black and gray dashed lines labeled by (VI) and (V) are additional parameter sweep lines. The red and blue dashed lines are phase boundaries, $p^A\approx 0.5$ and $p^C\approx 0.5$, respectively.
}
\label{Fig8}
\end{figure}
\section*{Appendix B: Computation of string topological order and scaling analysis}

The DWO can be calculated in the stabilizer formalism as $G(i_0,j_0)$ is written only by Pauli strings without imaginary factor $i$ and $G^2=1$. 
Each stabilizer $s^{\ell}(t)$ commutes or anti-commutes with $G$ at ${}^{\forall} t$, 
$Gs^{\ell}(t)=\alpha^{\ell}s^{\ell}(t)G$ with $\alpha^{\ell}=\pm 1$. 
The STO is reduced to a simple form
\begin{eqnarray}
O[({\rm DWO})^2] &=& 2^{N_{cs}}\mathrm{tr}[\rho(t)G(i_0,j_0)\rho(t)G(i_0,j_0)]
=\frac{2^{N_{cs}}}{2^{2N_R}}\mathrm{tr}\biggl[\prod^{N_R-1}_{\ell=0}(1+s^{\ell})G\prod^{N_R-1}_{k=0}(1+s^{k})G\biggl]\nonumber\\
&=&\frac{2^{N_{cs}}}{2^{2N_R}}\mathrm{tr}\biggl[\prod^{N_R-1}_{\ell=0}(1+s^{\ell})\prod^{N_R-1}_{k=0}(1+\alpha^{k}s^{k})\biggl]
=\frac{2^{N_{cs}}}{2^{2N_R}}\mathrm{tr}\biggl[\prod^{N_R-1}_{\ell=0}(1+s^{\ell})(1+\alpha^{\ell})\biggl]\nonumber\\
&=&\frac{2^{N_{cs}}}{2^{2N_R}}\mathrm{tr}\biggl[\prod^{N_R-1}_{\ell=0}(1+s^{\ell})\biggl]\prod^{N_R-1}_{k=0}(1+\alpha^{k})
=\frac{1}{2^{N_R}}\prod^{N_R-1}_{\ell=0}(1+\alpha^\ell),
\end{eqnarray}
where we have used $G^s(1+s^{\ell}(t))G^s=(1+G^ss^{\ell}(t)G^s)=(1+\alpha^{\ell}s^{\ell}(t))$.
For the ideal $Z_2\times Z_2$ SPT phase, due to $\alpha^{\ell}=1$ for ${}^{\forall} \ell$, $O[( {\rm DWO})^2]=1$ 
while for no $Z_2\times Z_2$ SPT phase, strictly $O[( {\rm DWO} )^2]=0$ due to due to $\alpha^{\ell}=-1$ for ${}^{\forall} \ell$. 

The observable $({\rm SG})^2$ in Eq.~(\ref{SGO}) is also calculated in a similar manner.

\section*{Appendix C: Additional data of the MoC simulation}
\begin{figure}[t]
\begin{center} 
\includegraphics[width=16cm]{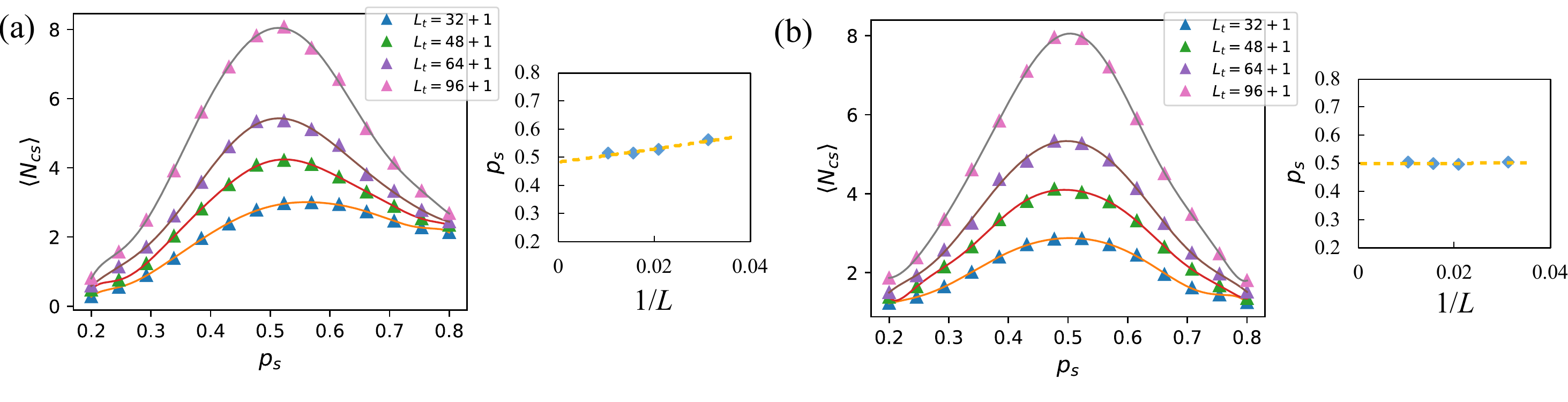}  
\end{center} 
\caption{Phase transition behaviors of $\langle N_{cs}\rangle$ for various system sizes. The solid lines are fitting lines. 
(a) The data for the case (IV), $p^A=p^{C}=p_s$ and 
(b) The data for the case (V), $p^A=1-p^C=p_s$.
Right small panels: System size dependence of $p_{sc}$ deduced by the peak of the fitting lines. We can extrapolate the critical probabilities for $1/L\to 0$. Here we used exponential fitting line.}
\label{Fig9}
\end{figure}

We further investigate the behavior of $\langle N_{cs}\rangle$ and the transition properties of mixed states in the MoC for additional parameter sweeps denoted by (V) and (VI) as shown in Fig.~\ref{Fig8}. 
For the case (V), we vary $p_s$ as defined by $p^A=p^{C}=p_s$ and for the case (VI), we vary $p_s$ defined by $p^A=1-p^C=p_s$.
The results of $\langle N_{cs}\rangle$ of the case (V) for different system sizes are displayed in Fig.~\ref{Fig9} (a). 
We find the value of $\langle N_{cs}\rangle$ clearly changes $0\to 2$ as increasing $p_{s}$. 
This implies that the mixed state transitions from the simple product phase to the Higgs=SPT phase in open boundary case. 
For all data, the system-size dependence emerges clearly and a peak of $\langle N_{cs}\rangle$ is located around $p_s\sim 0.5$. These peaks are signatures of a phase transition. 
The system-size dependence of $p_s$ of the peak is displayed in the right panel in Fig.~\ref{Fig9} (a). 
We can extrapolate the phase transition point $p_{sc}=0.483(5)$ for $L\to \infty$. 

Next, we show the results of $\langle N_{cs}\rangle$ of the case (VI) for different system sizes, displayed in Fig.~\ref{Fig9} (b). We observe that the value of $\langle N_{cs}\rangle$ clearly changes almost $1\to 1$ as increasing $p_s$. 
This implies the mixed state changes from the deconfinement phase to the SG phase in open boundary case. For all data, the system-size dependence emerges clearly and a peak of $\langle N_{cs}\rangle$ is located around $p_s\sim 0.5$. These peaks are signatures of a phase transition. 
The system-size dependence of $p_s$ of the peak point is displayed in the right panel in ~\ref{Fig9} (b). 
We can extrapolate the phase transition point $p_{sc}=0.499(8)$ for $L\to \infty$. 

These numerical results indicate that the probability point $(p^A,p^C)\sim(0.5,0.5)$ is a quadruple critical transition point.

\endwidetext



\begin{thebibliography}{99}

\bibitem{Li2018}
Y. Li, X. Chen, and M. P. A. Fisher, Phys. Rev. B {\bf 98}, 205136 (2018).

\bibitem{Skinner2019}
B. Skinner, J. Ruhman, and A. Nahum, Phys. Rev. X {\bf 9}, 031009 (2019).

\bibitem{Li2019}
Y. Li, X. Chen, and M. P. A. Fisher, Phys. Rev. B {\bf 100}, 134306 (2019).

\bibitem{Vasseur2019}
R. Vasseur, A. C. Potter, Y. -Z. You, and A. W. W. Ludwig, Phys. Rev. B {\bf 100}, 134203 (2019).

\bibitem{Chan2019}
A. Chan, R. M. Nandkishore, M. Pretko, and G. Smith, Phys. Rev. B {\bf 99}, 224307 (2019).

\bibitem{Szyniszewski2019}
M. Szyniszewski, A. Romito, and H. Schomerus, Phys. Rev. B {\bf 100}, 064204 (2019).

\bibitem{Choi2020}
S. Choi, Y. Bao, X.-L. Qi, and E. Altman, Phys. Rev. Lett. {\bf 125}, 030505 (2020).

\bibitem{Bao2020}
Y. Bao, S. Choi, and E. Altman, Phys. Rev. B {\bf 101}, 104301 (2020).

\bibitem{Jian2020}
C. -M. Jian, Y.-Z. You, R. Vasseur, and A. W. Ludwig, Phys. Rev. B {\bf 101}, 104302 (2020).

\bibitem{Zabalo2020}
A. Zabalo, M. J. Gullans, J. H. Wilson, S. Gopalakrishnan, D. A. Huse, and J. H. Pixley, Phys. Rev. B {\bf 101}, 060301 (2020).

\bibitem{Sang2021}
S. Sang and T. H. Hsieh, Phys. Rev. Res. {\bf 3}, 023200 (2021).

\bibitem{Sang2021_v2}
S. Sang, Y. Li, T. Zhou, X. Chen, T. H. Hsieh, and M. P. A. Fisher, PRX Quantum {\bf 2}, 030313 (2021).

\bibitem{Nahum2021}
A. Nahum, S. Roy, B. Skinner, and J. Ruhman, PRX Quantum {\bf 2}, 010352 (2021).

\bibitem{Sharma2022}
S. Sharma, X. Turkeshi, R. Fazio, and M. Dalmonte, SciPost Phys. Core {\bf 5}, 023 (2022).

\bibitem{Fisher2022_rev}
M. P. A. Fisher, V. Khemani, A. Nahum, and S. Vijay, arXiv:2207.14280.

\bibitem{Fuji2020}
Y. Fuji, and Y. Ashida, Phys. Rev. B {\bf 102}, 054302 (2020).

\bibitem{Lunt2020}
O. Lunt and A. Pal, Phys. Rev. Res. {\bf 2}, 043072 (2020).

\bibitem{Goto2020}
S. Goto, and I. Danshita, Phys. Rev. A {\bf 102}, 033316 (2020).

\bibitem{Tang2020}
Q. Tang, and W. Zhu, Phys. Rev. Research {\bf 2}, 013022 (2020).

\bibitem{Turkeshi2021}
X. Turkeshi, A. Biella, R. Fazio, M. Dalmonte, and M. Schiró, Phys. Rev. B {\bf 103}, 224210 (2021).

\bibitem{Kells2022}
G. Kells, D. Meidan, and A. Romito, arXiv:2112.09787 (2022).

\bibitem{Fleckenstein2022}
C. Fleckenstein, A. Zorzato, D. Varjas, E. J. Bergholtz, J. H. Bardarson, and A. Tiwari, Phys. Rev. Res. {\bf 4}, L032026 (2022).

\bibitem{KOH2022}
Y. Kuno, T. Orito, and I. Ichinose, Phys. Rev. B. {\bf 106}, 214304 (2022).

\bibitem{Lang2020}
N. Lang, and H. P. Büchler, Phys. Rev. B {\bf 102}, 094204 (2020).

\bibitem{Ippoliti2021}
M. Ippoliti, M. J. Gullans, S. Gopalakrishnan, D. A. Huse, and V. Khemani, Phys. Rev. X {\bf 11}, 011030 (2021).

\bibitem{Lavasani2021}
A. Lavasani, Y. Alavirad, and M. Barkeshli, Nat. Phys. {\bf 17}, 342 (2021).

\bibitem{Klocke2022}
K. Klocke and M. Buchhold, Phys. Rev. B {\bf 106}, 104307 (2022).

\bibitem{Lavasani2021_2}
A. Lavasani, Y. Alavirad, and M. Barkeshli, Phys. Rev. Lett. {\bf 127}, 235701 (2021).

\bibitem{Zeng2016}
B. Zeng and D. L. Zhou, EPL (Europhysics Letters) {\bf 113}, 56001 (2016).

\bibitem{Gottesman1997}
D. Gottesman, 
The Heisenberg representation of quantum computers, 
arXiv:9807006.

\bibitem{Aaronson2004}
S. Aaronson and D. Gottesman, Phys. Rev. A {\bf 70}, 052328 (2004).

\bibitem{Nielsen_Chuang}
M. A. Nielsen and I. L. Chuang, Quantum Computation and Quantum Information, by Michael A. Nielsen , Isaac L. Chuang, Cambridge, UK: Cambridge University Press, 2010 (2010).

\bibitem{Fradkin1979}
E. Fradkin and S. H. Shenker, Phys. Rev. D {\bf 19}, 3682 (1979).

\bibitem{Kogut1979}
J. B. Kogut, Rev. Mod. Phys. {\bf 51}, 659 (1979).

\bibitem{Verresen2022}
R. Verresen, U. Borla, A. Vishwanath, S. Moroz, and R. Thorngren, arxiv:2211.01376 (2022).

\bibitem{Borla2021}
U. Borla, R. Verresen, J. Shah, and S. Moroz, SciPost Phys. {\bf 10}, 148 (2021).

\bibitem{Gullans2020}
M. J. Gullans, and D. A. Huse, Phys. Rev. X {\bf 10}, 041020 (2020).

\bibitem{stab}
We assume that the stabilizer are Pauli operators without imaginary factor throughout this work.

\bibitem{Nahum2017}
A. Nahum, J. Ruhman, S. Vijay, and J. Haah, Phys. Rev X {\bf 7}, 031016 (2017).

\bibitem{Avella2013}
A. Avella, and F. Mancini, Strongly Correlated Systems: Numerical Methods, vol. 176, Springer, (2013).

\bibitem{Suzuki1976}
M. Suzuki, Progress of Theoretical Physics, {\bf 56}(5):1454–1469, 11 (1976).


\bibitem{Son2011}
W. Son, L. Amico, R. Fazio, A. Hamma, S. Pascazio, and V. Vedral, EPL (Europhysics Letters) {\bf 95}, 50001 (2011).

\bibitem{Son2012}
W. Son, L. Amico, and V. Vedral, Quantum Information Processing {\bf 11}, 1961 (2012).

\bibitem{Bahri2015}
Y. Bahri, R. Vosk, E. Altman, and A. Vishwanath, Nat. Commun. {\bf 6}, 7341 (2015).

\bibitem{Verrsen2017}
R. Verresen, R. Moessner, and F. Pollmann, Phys. Rev. B {\bf 96} 165124 (2017).


\bibitem{LGT_simu}
The way of imposing such a constraint of Gauss' law is also considered in study of quantum simulation of LGT in cold atoms in an optical lattice: 
E. Zohar, J. I. Cirac, and B. Reznik, Rep. Prog. Phys. {\bf 79}, 014401 (2015); K. Kasamatsu, I. Ichinose, and T. Matsui, Phys. Rev. Lett. {\bf 111}, 115303 (2013); Y. Kuno, K. Kasamatsu, Y. Takahashi, I. Ichinose, and T. Matsui, New J. Phys. {\bf 17}, 063005 (2015); Y. Kuno, S. Sakane, K. Kasamatsu, I. Ichinose, and T. Matsui, Phys. Rev. D {\bf 95}, 094507 (2017).


\bibitem{deconfinement_stab}
Since, for $K_1 \to \infty$, $J_2\to \infty$ and $K_2,J_1=0$, 
$L$ stabilizers from $K_1$-term and $L$ stabilizers from $J_2$-term stabilize the state and then, the one redundant degree of freedom exists as $L_t-(L+L)=1$, producing two-fold degeneracy of the ground state under open boundary conditions.


\bibitem{QI_text}
B. Zeng, X. Chen, D. Zhou, and X. -G. Wen, “Quantum Information Meets Quantum Matter,” Quantum Science and Technology (2019).

\bibitem{Ncs_fit}
The data point $\langle N_{cs}\rangle$ is fit by 10-th order polynomial function.

\bibitem{exp_fit}
This choice was determined by $R^2$ values compared with various fitting function forms.

\bibitem{Takashima2005}
S. Takashima, I. Ichinose, and T. Matsui, Phys. Rev. B {\bf 72}, 075112 (2005).

\end{thebibliography}
\end{document}